\documentclass[secnumarabic, twocolumn, graphics,floatfix,nofootinbib,
tightenlines,nobibnotes,aps,prb,10pt]{revtex4}

\usepackage{color}
\usepackage{graphicx}
\usepackage{placeins}
\usepackage{amsmath}
\usepackage{nicefrac}

\newcommand{\FeAs}{Fe$_2$As$_2$}

\newcommand{\aaxis}{$a$}
\newcommand{\baxis}{$b$}
\newcommand{\caxis}{$c$}
\newcommand{\abplane}{$ab$}

\begin{document}

%===============================================================================
%  TITLE
%===============================================================================

\title{Ab-initio perspective on structural and electronic properties of 
iron-based superconductors}

%===============================================================================
%  AUTHORS
%===============================================================================

\author{Daniel Guterding}
\affiliation{Institut f\"ur Theoretische Physik, Goethe-Universit\"at Frankfurt,
Max-von-Laue-Stra{\ss}e 1, 60438 Frankfurt am Main, Germany}

\author{Steffen Backes}
\affiliation{Institut f\"ur Theoretische Physik, Goethe-Universit\"at Frankfurt,
Max-von-Laue-Stra{\ss}e 1, 60438 Frankfurt am Main, Germany}

\author{Milan Tomi{\'c}}
\affiliation{Institut f\"ur Theoretische Physik, Goethe-Universit\"at Frankfurt,
Max-von-Laue-Stra{\ss}e 1, 60438 Frankfurt am Main, Germany}

\author{Harald O. Jeschke}
\affiliation{Institut f\"ur Theoretische Physik, Goethe-Universit\"at Frankfurt,
Max-von-Laue-Stra{\ss}e 1, 60438 Frankfurt am Main, Germany}

\author{Roser Valent\'\i}
\affiliation{Institut f\"ur Theoretische Physik, Goethe-Universit\"at Frankfurt,
Max-von-Laue-Stra{\ss}e 1, 60438 Frankfurt am Main, Germany}

%===============================================================================
%  ABSTRACT
%===============================================================================

\begin{abstract}
The discovery of iron pnictides and iron chalcogenides as a new class of 
unconventional superconductors in 2008 has generated an enourmous amount of 
experimental and theoretical work that identifies these materials as correlated 
metals with multiorbital physics, where magnetism, nematicity and 
superconductivity are competing phases that appear as a function of pressure and 
doping. A microscopic understanding of the appearance of these phases is crucial 
in order to determine the nature of superconductivity in these systems. Here we 
review our recent theoretical efforts to describe and understand from first 
principles the properties of iron pnictides and chalcogenides with special focus 
on (i) pressure dependence, (ii) effects of electronic correlation and (iii) 
origin of magnetism and superconductivity.
\end{abstract}

\maketitle

%===============================================================================
%  INTRODUCTION
%===============================================================================

\section{Introduction}
\label{sec:introduction}
The discovery of iron-based superconductivity in LaFeAsO$_{1-x}$F$_x$ with a 
$T_c = 26~\mathrm{K}$~\cite{Kamihara2008} created a new field of research and 
incited intense experimental and theoretical work in this area. Here we review 
our present theoretical knowledge of the microscopic behavior of these 
materials. In particular,  we elucidate
 via first principles investigations the influence
of  pressure, correlations and, to a less extent, doping
 on the electronic, magnetic and superconductor 
properties of these materials.
 For our analysis we consider a combination of
 density functional theory (DFT), 
dynamical mean-field theory (DMFT) and spin fluctuation theory.

%===============================================================================
%  OVERVIEW OF FE BASED SCs
%===============================================================================

\begin{figure}[tb]
\includegraphics[width=0.8\linewidth]{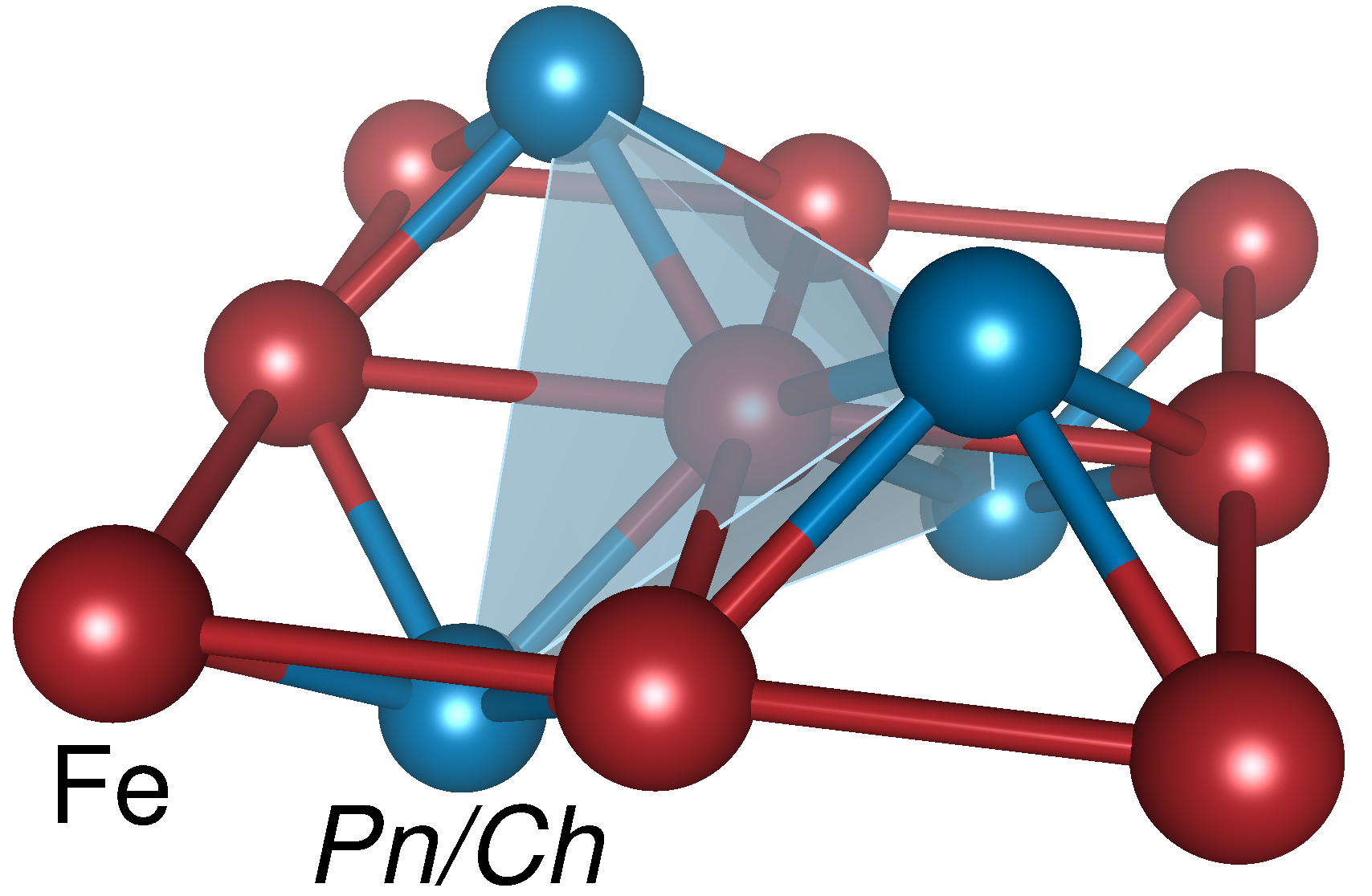}
\caption{Structural building block of iron pnictide and chalcogenide 
superconductors. $Pn$ stands for the pnictogen phosphorous or arsenic, $Ch$ for 
the chalcogen sulphur, selenium or tellurium.}
\label{fig:structure}
\end{figure}

{\it Crystal structure.-} The basic building block of all iron-based 
superconductors is a two-dimensional square lattice of iron atoms, tetrahedrally 
coordinated by pnictogen or chalcogen atoms (Fig.~\ref{fig:structure}). The 
resulting structure is a tri-layer, where the iron layer is sandwiched between 
two layers of pnictogen or chalcogen atoms. The C$_4$ symmetry of the iron 
lattice translates into the tetragonal symmetry of the overall crystal lattice, 
unless broken by magnetic ordering.  Trilayers can be stacked in different 
fashions. Between layers, there can be (i) no ions (FeSe as representative of the 11 
family of chalcogenides), (ii) alkali ions (LiFeAs as typical member of the so called 
111 family of pnictides), (iii) alkaline earth ions (Ca{\FeAs} is the lightest 
representative of the 122 family), (iv) rare earth oxides (LaFeAsO as prototype of 
the 1111 family) and (v) even perovskites and organic molecules. The crystal 
structure of the various families is referred to by the stoichiometry of the 
corresponding formula unit.  In this review we consider iron-based 
superconductors with unit cells containing either one or two tri-layers.  We 
especially focus on the 122 family with two trilayers in the unit cell, which 
belongs to the ThCr$_2$Si$_2$ crystal structure type~\cite{Hoffmann1985}. A well 
known instability of these structures is a collapse along the 
crystallographic \caxis-axis, when bonds are formed between the pnictogen 
sublayers of adjacent trilayers.

{\it Electronic structure and pressure effects.-} Iron-based superconductors are 
$3d$ metals with iron in a nominal $+2$ oxidation state (Fe $3d^6$). In the 
(imperfect) tetrahedral environment of pnictogen/chalcogen, iron $3d$ orbitals 
hybridize with the  pnictogen/chalcogen $p$ orbitals and split approximately 
into the doubly degenerate $e_g$ and triply degenerate $t_{2g}$ orbitals. The 
physics of iron-based superconductors is fundamentally multi-orbital involving 
all five $3d$ orbitals and to a lesser extent the pnictogen/chalcogen $p$ 
orbitals. The physics at the Fermi level is dominated by the $t_{2g}$ orbitals, 
with a subdominant role played by $e_g$ and $p$ orbitals.

When pressure is applied to these systems, it couples to bond lengths and 
angles, forcing new equilibrium positions of all atoms within a changed unit 
cell. Therefore, pressure is a very direct way of modifying the crystal
structure,  
electronic properties and magnetism~\cite{Opahle2009}.  The 122 family of iron pnictides has been the focus of 
many experimental and theoretical pressure studies. Of particular interest are 
Ca{\FeAs}, where As $p$-As $p$ bonds form easily along $c$ and lead to a structural
collapse, and Ba{\FeAs}, where the larger Ba atom suppresses interlayer bonding. 
At ambient pressure and a temperature of 172 K, Ca{\FeAs} undergoes a sharp first 
order transition from the paramagnetic tetragonal phase into the orthorhombic 
antiferromagnetically ordered phase~\cite{Ni2008, Goldman2008, Wu2008, 
Ronning2008, Canfield2009, Alzamora2011}. This transition can be suppressed by 
the application of pressure~\cite{Yu2009, Canfield2009, Lee2009, Kreyssig2008, 
Torikachvili2008, Goldman2009} and the paramagnetic collapsed tetragonal phase can 
be observed at around 0.4 GPa. These systems show high sensitivity to the 
hydrostaticity of the applied pressure~\cite{Prokes2010, Yu2009, Canfield2009, 
Torikachvili2009}. Interestingly, under good hydrostatic conditions Ca{\FeAs} 
does not show any signal of superconductivity~\cite{Yu2009}, which is however the case 
when a non-hydrostatic component is present. In addition, application of 
purely uniaxial pressure along the crystallographic \caxis-axis reduces the 
pressure at which the collapsed tetragonal occurs by an order of magnitude to 
0.06 GPa~\cite{Prokes2010}.

Ba{\FeAs} undergoes a transition from the tetragonal paramagnetic into the 
orthorhombic antiferromagnetic phase at ambient pressure and a temperature of 
140 K~\cite{Rotter2008}. Application of pressure within a certain range produces 
a superconducting dome~\cite{Colombier2009, Alireza2009, Paglione2010}, beyond 
which magnetic order is suppressed and a tetragonal phase emerges, followed by 
the collapsed tetragonal phase at even higher pressures~\cite{Uhoya2010, 
Mittal2011}. The situation in Ba{\FeAs} is complicated by phase coexistence. The 
tetragonal signature can be observed already at 6 GPa by neutron diffraction 
~\cite{Kimber2009} although Ba{\FeAs} stays magnetically ordered up to 10 
GPa~\cite{Mittal2011}.  Sensitivity to non-hydrostaticity along the \caxis-axis 
has been also reported~\cite{Uhoya2010}.

Various strain conditions on the \abplane-plane have been also investigated and 
are presently still a subject of intensive discussion.  Experimentally, in-plane 
application of tensile strain is used to detwin  samples~\cite{Chu2010, 
Tanatar2010, Kuo2011, Liang2011, Dhital2012, Blomberg2012, Chu2012} in order to 
provide a better insight into the anisotropic properties, in particular, in 
relation to the so-called nematic phase ~\cite{Fang2008, Xu2008, Fernandes2010}. 
In addition, it has been shown that in-plane strain has a significant impact on 
the magnetic properties of Ba{\FeAs}~\cite{Dhital2012, Kuo2012}.
In  section \ref{sec:pressure}
 we present our simulations on 
pressure effects in the 122 systems.

{\it Correlation effects.-} An important aspect of iron-based superconductors is 
the role of electronic correlations in determining the behavior of these 
systems. The investigation of correlation effects in these materials has been a 
subject of intensive research since their discovery. 
The observation of significant  
band renormalizations and mass enhancements in 
 optical spectroscopy~\cite{Qazilbash2009}, photoemission
spectroscopy~\cite{Sato2009, Yoshida2011, Yoshida2014} and quantum oscillation
experiments~\cite{Coldea2008,Terashima2010,Carrington2011,Putzke2012,Terashima2013A}
or the detection of an incoherent bad metal to coherent Fermi liquid
 phase transition at low
temperatures~\cite{Hardy2013,Boehmer_thesis2014}
are experimental examples that
clearly set these systems as correlated metals. However, the true nature
of these materials is still a subject of debate; namely whether 
these materials are
on the verge of being Mott insulators or, alternatively, they behave
as Hund's metals.

While 
the metallic nature of these materials has made a 
 DFT-based description enormously successful~\cite{Mazin2009}, there are many 
aspects which are less well captured within DFT like band renormalizations
and mass enhancements:
DFT bands and Fermi surfaces differ
 quantitatively and sometimes qualitatively from experimental observations. 
Also, the absence of quantum fluctuations in the magnetic DFT description has 
some serious consequences. Therefore, an improved treatment of electronic 
correlations has been discussed to be important for quantitative comparisons 
with experiment.

A method that has proven quite successful in capturing the essential features of 
electronic correlations in iron-based superconductors is the combination of 
density functional theory with dynamical mean-field theory 
(DFT+DMFT)~\cite{Yin2011,Aichhorn2010,Hansmann2010,Aichhorn2011, 
Ferber2012,Ferber2012a,Werner2012}.  It treats both itinerant and localized 
properties of the electrons on equal footing. Many studies have dealt with the 
experimentally observed effects of correlation like large masses enhancements or 
possible non-Fermi liquid 
behavior~\cite{Haule2009,Aichhorn2010,Yin2011,de'Medici2011A,Yu2011, 
Ferber2012,Ferber2012a}.  The physics of iron-based
superconductors  is controlled by all Fe 
$3d$ orbitals, leading to a multiple orbital problem crucially influenced by the Hund's 
coupling $J_H$~\cite{Haule2009,Aichhorn2009,Liebsch2011,de'Medici2011A, 
Yu2011,Ferber2012,Ferber2012a,Georges2013}. However, the relative importance and 
the role of $J_H$ versus the on-site Coulomb repulsion $U$ is an ongoing debate 
in the interpretation of the correlated nature of Fe-pnictides and 
Fe-chalcogenides~\cite{Haule2009,Yin2011,de'Medici2011A, 
Yao2011,Georges2013,Yu2013,de'Medici2014,Fanfarillo2015}. An important 
insight has been gained in several studies by recognizing that depending on the 
electronic filling, the Hund's coupling $J_H$ can, on the one hand, render a 
moderately correlated system even more correlated and push it into a bad metal 
regime, while, on the other hand, it can also reestablish a metallic behavior, 
albeit orbital selective, in a strongly correlated 
system~\cite{Georges2013,Fanfarillo2015}. 

For the investigation of typical correlation effects such as band and effective 
mass renormalizations, as well as  Hubbard satellites, in section \ref{sec:correlations} we 
 review
LiFeAs, LiFeP, LaFePO, Ca{\FeAs} and the hole-doped $A$Fe$_2$As$_2$ ($A={\rm 
K}$, Rb, Cs) end members of the 122 iron pnictide series in order to allow a 
comparative analysis of different degrees of correlation and their consequences 
on the properties of Fe-based superconductors. The latter family of systems is 
ideal for investigating the effects of correlation versus {\it negative} 
pressure, as the unit cell expands along the series. The removal of one electron 
per formula unit by substitution of Ba by K in Ba{\FeAs} is accompanied by a 
complete suppression of any structural or magnetic phase 
transition~\cite{Paglione2010,Boehmer_thesis2014} and by the emergence of 
superconductivity at low temperatures~\cite{Tafti2013}. This behavior is quite 
generic in all hole-doped end members 
$A$Fe$_2$As$_2$~\cite{Kihou2010,Dong2010,Eilers_thesis2014,Zhang2015,Hong2013}.

There is also experimental evidence that the parent compound 
Ba$_{1-x}$K$_x$Fe$_2$As$_2$ undergoes a coherence-incoherence 
transition~\cite{Werner2012,Hardy2013,Liu2014} as a function of temperature, 
probably caused by a strong increase in correlations, since the system is pushed 
closer to half filling~\cite{Popovich2010,Mu2009,Terashima2010}. Experimental 
determination of the Sommerfeld coefficient seems to indicate that these 
hole-doped end systems are one the most strongly correlated known 122 
iron-pnictide superconductors~\cite{Hardy2013,Eilers_thesis2014}, which is also 
indicated by multiple theoretical investigations on 
K{\FeAs}~\cite{Yin2011,Hardy2013,Skornyakov2014,de'Medici2014}. Along the doping 
series from Ba{\FeAs} to K{\FeAs} the Sommerfeld coefficient increases by more 
than an order of magnitude~\cite{Hardy2013,Boehmer_thesis2014,Storey2013} and 
further continues to increase as K is substituted by atoms with larger atomic 
radius like the isovalent Rb and Cs~\cite{Shermadini2010,Zhang2015}. 

{\it Superconductivity.-}
It was realized early on that superconductivity in iron-based materials is 
unconventional and probably mediated by spin-fluctuations~\cite{Mazin2008A}. 
Although this is not ultimately settled, research in this area has become 
relatively mature and a number of reviews on the topic have 
appeared~\cite{Hirschfeld2011,Scalapino2012,Dagotto2013,Shibauchi2014,
Hosono2015a}. It is widely believed that the strong orbital differentiation 
and almost two-dimensional electronic structure are the key features of 
iron-based superconductors. The  questions remaining to be answered are (i) 
whether a unified model of iron-based superconductors exists and (ii) how the 
superconducting transition temperature can be optimized. In 
section \ref{sec:superconductivity} we  
discuss our present understanding of superconductivity by reviewing the 
superconducting behavior of a few families of iron-based superconductors.
 
\section{Methods}
\label{sec:methods}
Our investigations are based on first principles calculations
combining density functional theory,
dynamical mean-field theory and spin fluctuation theory.
In this section we present the details of the three approaches.

{\it DFT calculations.-}
For the density functional theory calculations we use the all-electron full-potential local orbital 
(FPLO)\cite{Koepernik1999} code in the generalized gradient approximation 
(GGA)\cite{Perdew1996A}, as well as the \textsc{WIEN2k}~\cite{Blaha01} 
implementation of the full-potential linear augmented plane wave (FLAPW) method 
in both GGA and the local density approximation (LDA) and we also employed the
 Vienna ab initio simulations package (VASP)~\cite{Kresse1993,Kresse1996} with the 
projector augmented wave (PAW) basis~\cite{Bloechl1994}. All of our structural 
relaxations presented were performed under constant stress using the Fast Inertial 
Relaxation Engine (FIRE)~\cite{Bitzek2006}, with a 
modified relaxation algorithm~\cite{Tomic2013}.

{\it LDA+DMFT calculations.-}
We combine the DFT method with dynamical mean-field theory 
(DMFT) to include electronic correlation effects beyond the local density 
approximation. In the DMFT approximation one assumes
 that the coordination number $Z$ ( number of nearest neighbours)
 is large, so that non-local fluctuations are small
because they tend to be averaged out for large $Z$. As has been shown~\cite{MetznerVollhardt89,MuellerHartmann89},
in the limit of $Z\rightarrow \infty$, this approximation is exact and the self-energy becomes
a local quantity
\begin{align}
 \Sigma_{ij}(\omega) \rightarrow \delta_{ij}\Sigma_{ii}(\omega),
\end{align}
where $i,j$ label the atomic sites, and correspondingly, its Fourier transform 
is momentum independent. In this limit, the self-energy can be obtained by a self-consistent
solution of an effective Anderson impurity model~\cite{GeorgesKotliar96}.
With this, the interacting Green's function can be written as 
\begin{equation}
 G(k,\omega)
= \left[ \omega + i\delta +\mu -\epsilon_k -\Sigma(\omega) \right]^{-1},
\end{equation}
where $\delta>0$ is a small convergence parameter. The dispersion $\epsilon_k$
is given by the non-interacting system, which is approximatively given
(minus a doublecounting term) by the DFT result.

We implemented our own version of the LDA+DMFT cycle (see 
Ref.~\onlinecite{Backes2014} for a more detailed explanation), in combination with the 
continuous-time quantum Monte Carlo method in the hybridization 
expansion~\cite{Werner06} as implemented in the ALPS~\cite{ALPS11,Gull11a} 
project for solving the effective impurity model.
For the DFT calculations we used the \textsc{WIEN2k}~\cite{Blaha01} 
implementation of the FLAPW method 
in the local density approximation. A local orbital basis was obtained by a 
projection of the Bloch wave functions to the localized Fe $3d$ orbitals, using 
our implementation of the projection described in 
Refs.~\onlinecite{Aichhorn2009,Ferber2014}. 
The interaction parameters were used in the
definition of the Slater integrals~\cite{Liechtenstein95} $F^k$ with
$U=F^0$ and $J_H=(F^2+F^4)/14$. 
Observables like the effective masses can be directly calculated from the impurity self-energy via
\begin{equation}
    \frac{m^\ast}{m_\mathrm{LDA}} = 1 - \left.\frac{\partial\mathrm{Im}\Sigma(i\omega)}{\partial i\omega}\right|_{\omega\rightarrow 0^+},
    \label{eq:meff}
\end{equation}
with $i\omega$ on the Matsubara axis.
The continuation of the Monte Carlo data to the real axis was done by stochastic
analytic continuation~\cite{Beach2004}.

{\it RPA spin-fluctuation calculations.-}
Soon after the discovery of iron-based superconductors it was suggested that 
superconductivity in these materials might be unconventional and mediated by 
antiferromagnetic spin-fluctuations~\cite{Mazin2008A}. Subsequently, several 
groups developed methods of calculating the symmetry of the superconducting 
order parameter and the pairing strength based on the electronic bandstructure 
and an interaction term (for a review see Ref.~\onlinecite{Hirschfeld2011}). One 
of those methods is the random phase approximation (RPA) approach to the 
multi-orbital Hubbard model~\cite{Graser2009,Kreisel2013}, which we briefly 
review here.

In this method, the Hamiltonian consists of a kinetic part $H_0$ and an 
interaction term $H_\mathrm{int}$. The kinetic part is usually a Wannier 
representation of the DFT bandstructure. The crystallographic unit cell 
contains two iron atoms, contributing $2 \times 5$ $d$-orbitals close to the 
Fermi level, and two pnictogen or chalcogen atoms, contributing $2 \times 3$ 
$p$-orbitals close to the Fermi level. In consequence, a good representation of 
the DFT bands can be obtained with a 16-orbital tight binding model. Such a 
model of the two-iron Brillouin zone can be approximately unfolded to an 
8-orbital model of the effective one-iron Brillouin zone by using the glide 
reflection symmetry of the unit cell. To handle this we have developed a 
generalized unfolding method relying on induced representations 
of space groups~\cite{Tomic2014}. For the interaction term 
$H_\mathrm{int}$ the multi-orbital Hubbard interaction is used on the iron site. 
Here, $\sigma$ represents the spin, $n_{i l \sigma} = c^\dagger_{i l \sigma}c^{\,}_{i 
l \sigma}$ and $n_{il} = n_{i l \uparrow} + n_{i l \downarrow}$. The indices $l$
denote the atomic orbitals. 
\begin{equation}
\begin{split}
H =& H_0 + H_\mathrm{int} \\ =& \sum\limits_{l_1, l_2 \sigma} \sum\limits_{<ij>}
t_{ij}^{l_1 l_2} c^\dagger_{i l_1 \sigma} c^{\,}_{j l_2 \sigma} + U \sum\limits_{i,l} 
n_{i l \uparrow} n_{i l \downarrow}\\
& + \frac{V}{2} \sum\limits_{i,l_1,l_2 \neq l_1} n_{i l_1} n_{i l_2}
- \frac{J}{2}\sum\limits_{i,l_1,l_2\neq l_1}
\vec S_{i l_1} \cdot \vec S_{i l_2}\\
&+ \frac{J^\prime}{2} \sum \limits_{i,l_1,l_2 \neq l_1, \sigma} 
c^\dagger_{i l_1 \sigma} c^\dagger_{i l_1 \bar \sigma}
c^{\,}_{i l_2 \bar \sigma} c^{\,}_{i l_2 \sigma}
\end{split}
\label{eq:hamiltonian}
\end{equation}

Subsequently, the non-interacting static susceptibility $\chi^0$ is calculated,
\begin{equation}
\begin{split}
\chi^0_{{l_1} {l_2} {l_3} {l_4}} (\vec q) = - \frac{1}{N} \sum 
\limits_{\vec k, \mu, \nu} & a_\mu^{l_4} (\vec k) a_\mu^{l_2 *}
(\vec k) a_\nu^{l_1} (\vec k + \vec q) a_\nu^{l_3 *} (\vec k + \vec q) \\
 &\displaystyle \times \frac{f(E_\nu (\vec k + \vec q)) - 
 f(E_\mu (\vec k))}{E_\nu (\vec k + \vec q) - E_\mu (\vec k)}
\end{split}
\label{eq:nonintsuscep}
\end{equation}
where matrix elements $a^l_{\mu} (\vec k)$ resulting from the diagonalization 
of the initial Hamiltonian $H_0$ connect orbital and band-space denoted by 
indices $l$ and $\mu$ respectively. The $E_\mu$ are the eigenvalues of $H_0$ and 
$f(E)$ is the Fermi function. $N$ is the number of sites in the unit cell. 
Temperature enters the 
calculation through the Fermi functions. 

In the next step, the static spin- and orbital-susceptibilities 
($\chi^{s,\mathrm{RPA}}$ and $\chi^{c,\mathrm{RPA}}$) are constructed in an RPA 
framework. The structure of the interaction matrices $U^s$ and $U^c$ can be
inferred from 
Eq.~\ref{eq:hamiltonian} (see Ref.~\onlinecite{Graser2009}).
\begin{subequations}
\begin{align}
\Big[ \chi^{s,\mathrm{RPA}}_{{l_1} {l_2} {l_3} {l_4}} ({\vec q}) \Big]^{-1} =& 
\Big[ \chi^{0}_{{l_1} {l_2} {l_3} {l_4}} ({\vec q}) \Big]^{-1} 
- U^s_{{l_1} {l_2} {l_3} {l_4}} \\
\Big[ \chi^{c,\mathrm{RPA}}_{{l_1} {l_2} {l_3} {l_4}} ({\vec q}) \Big]^{-1} =& 
\Big[ \chi^{0}_{{l_1} {l_2} {l_3} {l_4}} ({\vec q}) \Big]^{-1} 
+ U^c_{{l_1} {l_2} {l_3} {l_4}}
\end{align}
\label{eq:rpasuscep}
\end{subequations}

The pairing vertex in orbital space for the spin-singlet channel can be 
calculated using the fluctuation exchange approximation, which uses the 
previously calculated RPA susceptibilities. In the pairing vertex momenta $\vec 
k$ and $\vec k^\prime$ are restricted to the Fermi surface. 
\begin{equation}
\begin{split}
\Gamma_{{l_1} {l_2} {l_3} {l_4}} (\vec k, \vec k^\prime) =
&\left[\frac{3}{2} {U^s} \chi^{s,\mathrm{RPA}} (\vec k - \vec k^\prime) {U^s}
+ \frac{1}{2}U^s \right.\\
- &\left.\frac{1}{2} {U}^c \chi^{c,\mathrm{RPA}} ({\vec k}
- {\vec k^\prime}) {U^c} + \frac{1}{2}{U^c} \right]_{{l_1} {l_2} {l_3} {l_4}}
\end{split}
\label{eq:pairingvertexorbitalspace}
\end{equation}

The pairing vertex in orbital space is transformed into band space using the
matrix elements $a^l_{\mu} (\vec k)$.
\begin{equation}
\begin{split}
\Gamma_{\mu \nu} (\vec k, \vec k^\prime) = \mathrm{Re} 
\sum\limits_{l_1 l_2 l_3 l_4} &a_\mu^{l_1,*} (\vec k) a_\mu^{l_4,*} (-\vec k)
[\Gamma_{{l_1} {l_2} {l_3} {l_4}} (\vec k, \vec k^\prime)]\\ 
\times & a_\nu^{l_2} (\vec k^\prime) a_\nu^{l_3} (-\vec k^\prime)
\end{split}
\label{eq:pairingvertexbandspace}
\end{equation} 

Finally, one solves the linearized gap equation
\begin{equation}
- \sum \limits_\nu \oint_{C_\nu} \frac{dk^\prime_\parallel}{2\pi} 
\frac{1}{2\pi \, v_F (\vec k^\prime)} \left[ \Gamma_{\mu\nu}
(\vec k, \vec k^\prime) \right]
g_i (\vec k^\prime) = \lambda_i g_i (\vec k)
\label{eq:gapequation}
\end{equation}
 by performing an 
eigendecomposition on the kernel and obtains the dimensionless pairing strength 
$\lambda_i$ and the symmetry function $g_i (\vec k)$. The integration runs over
the discretized Fermi surface and $v_F (\vec k)$ is the magnitude of the Fermi
velocity.

This method of calculating the superconducting order parameter and pairing 
strength is appropriate when the pairing interaction quickly drops as a function 
of frequency, i.e. only Fermi surface pairing plays a role. In situations where
bands away from the Fermi level are expected to contribute to the pairing, 
a frequency-dependent formulation has to be employed.

\section{Simulating the effects of pressure} 
\label{sec:pressure}
\begin{figure}[tb]
\includegraphics[width=0.49\textwidth]{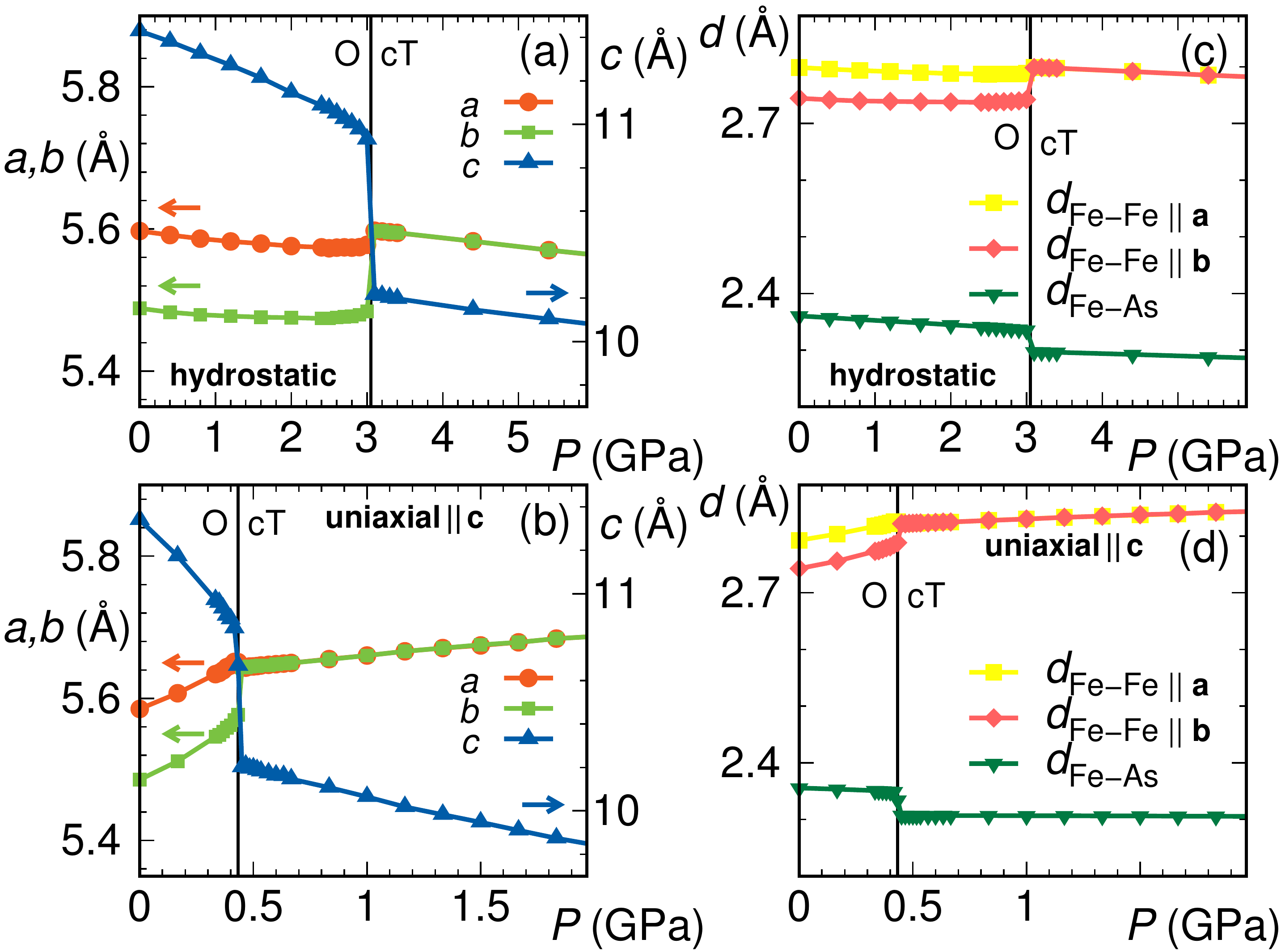}
\caption{Structure parameters of Ca{\FeAs} under application of hydrostatic (top 
row) and uniaxial pressure (bottom row). Shown are {\it ab-initio} calculated 
lattice parameters (a,b), and Fe-Fe and Fe-As bond lengths (c,d)
within DFT (GGA).
Reprinted with permission from Ref.~\onlinecite{Tomic2012}.} 
\label{fig:ca_press_struct}
\end{figure}

We review here Ca{\FeAs} and Ba{\FeAs} as representative examples of the 122 
family regarding pressure effects and the corresponding theoretical simulations 
with special emphasis on the origin and consequences of the appearance of a 
collapsed tetragonal phase.

\begin{figure*}[t]
\begin{center}
\includegraphics[width=0.8\textwidth]{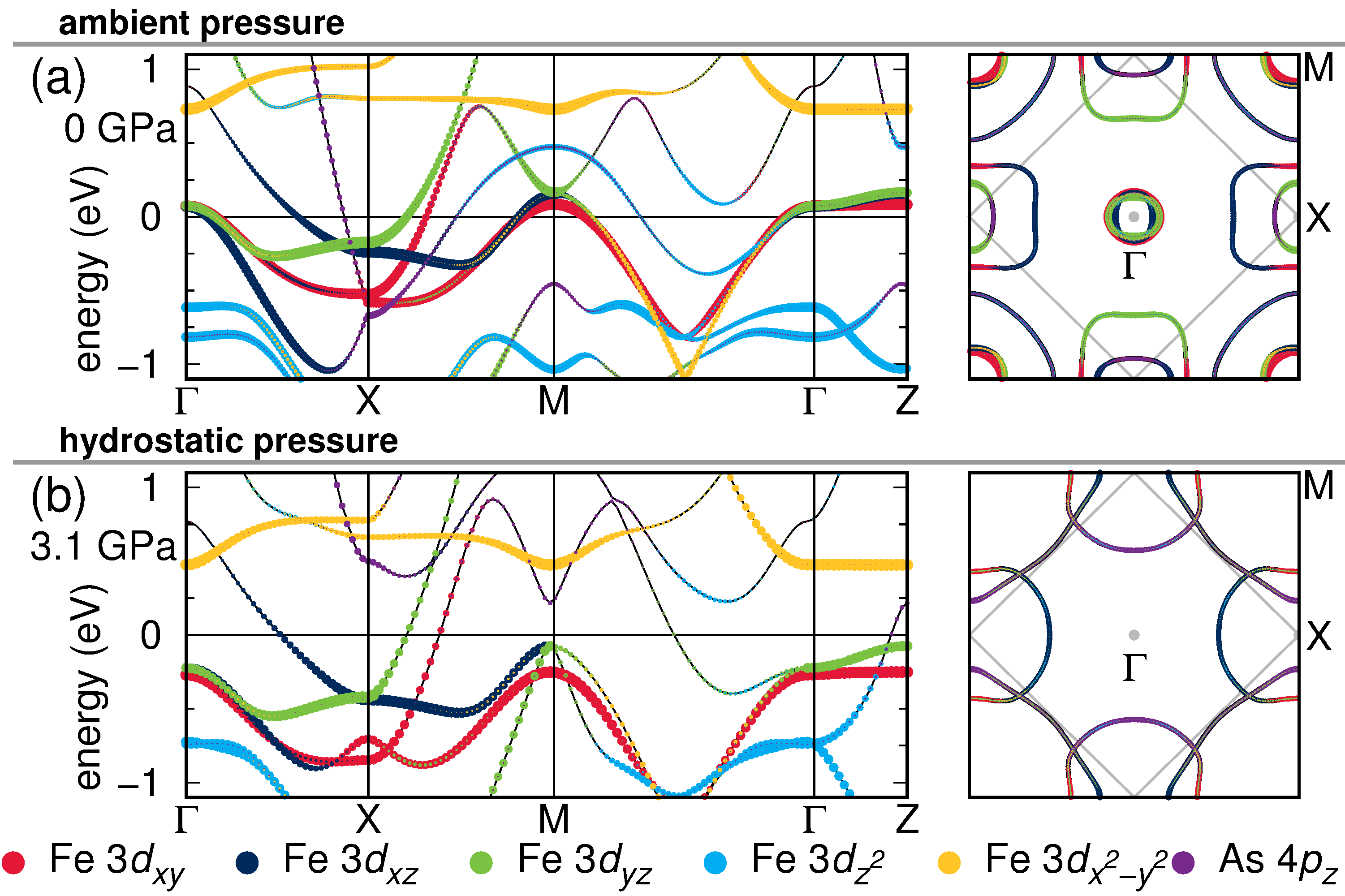}
\end{center}
\caption{Electronic structure of Ca\FeAs~under hydrostatic pressure calculated 
from DFT (GGA). The band structure and the Fermi surface are shown in the large 
Brillouin zone corresponding to the 1Fe unit cell. Reprinted with permission 
from Ref.~\onlinecite{Tomic2012}.}\label{fig:ca122_el_str}
\end{figure*}

Under hydrostatic pressure the \caxis-axis of Ca{\FeAs} undergoes a contraction 
at a more rapid rate than the \abplane-plane [see 
Fig.~\ref{fig:ca_press_struct}(a)]~\cite{Tomic2012}. This is to be expected as 
122 compounds don't have any chemical bonds oriented along the \caxis-axis, and 
thus all structural deformations along the \caxis-axis are bond-bending and low 
in energy. This is not the case for the \abplane-plane, where Fe-Fe bonds orient 
along the \aaxis~and \baxis-axes. At around 3.1~GPa the distance between the two 
adjacent tri-layers becomes small enough that interlayer As $p$-As $p$ bonds 
form and the \caxis-axis undergoes a sharp collapse of 6.5\%. At the same time 
the \baxis-axis abruptly expands to assume the same length as the \aaxis-axis. 
In total the unit cell volume drops by 4.3\% [see 
Fig.~\ref{fig:ca_press_struct}(c)] and the symmetry becomes tetragonal in 
absence of magnetism. The ratio $c/a_t=2.58$ [see 
Fig.~\ref{fig:ca_press_struct}(c)], with $a_t=a/\sqrt{2}$, of the tetragonal 
cell indicates the structural collapse. This is in good qualitative agreement 
with experimental observations~\cite{Kreyssig2008}. The overestimation of the 
transition pressure, which is experimentally determined to be around 0.5~GPa, is 
also observed in other theoretical studies~\cite{Colonna2011, Zhang2009}, and is 
the consequence of the sharp first order nature of the transition. The estimated 
bulk modulus at  ambient pressure is $70\pm 3$ GPa, in good agreement with 
the experimentally observed value~\cite{Joergensen2010}, and it increases to 
$105\pm 2$ GPa at the transition to the collapsed tetragonal phase. 

The Fe-As bond lengths undergo a contraction in the entire pressure range, with 
a sharp drop at the transition pressure. In terms of a local moment picture, 
this leads to an increased crystal field splitting, and a subsequent suppression 
of the iron magnetic moments, which is consistent with the observed lack of 
magnetic order in the collapsed tetragonal phase. 

In terms of the electronic structure, the increase of pressure pushes the 
$t_{2g}$ band manifold towards lower energies as seen in 
Fig.~\ref{fig:ca122_el_str}(a), resulting in reduced contributions of $d_{xy}$, 
$d_{xz}$ and $d_{yz}$ orbitals to the Fermi level density of states. At the same 
time, this means that the hole Fermi surface pockets around the $\Gamma$ point 
become smaller, while the electron pockets around the $\bar{X}$ point become 
larger [see Fig.~\ref{fig:ca122_el_str}(b)]. The label $\bar{X}$ denotes the 
$X$-point of the Brillouin zone commensurate with the unit cell of the iron 
sublattice (the so-called 1Fe unit cell). An immediate consequence is the 
worsening of Fermi surface nesting, which leads to a weakening of the spin 
density wave state. Going deeper into the collapsed tetragonal phase, the three 
dimensionality of the Fermi surface becomes more pronounced as the tri-layers 
come closer.

Because of the inability to perform the ARPES experiments under pressure, a 
direct experimental observation of the aforementioned Fermi surface effects is 
not straightforward. However, due to the sharp first order nature of the phase 
transition in Ca{\FeAs}, it was shown that stabilization of the collapsed 
tetragonal phase is possible through postgrowth annealing and quenching of the 
samples~\cite{Ran2011, Ran2012, Gati2012}. Through the rapid quenching from high 
temperature, Ca{\FeAs} samples were essentially frozen in the metastable 
internally strained state, such that they do not revert to the orthorhombic 
phase upon cooling. Band dispersions observed with ARPES were shown to be in 
excellent agreement with the theoretically predicted picture~\cite{Dhaka2014}.

\begin{figure}[tb]
\includegraphics[width=0.45\textwidth]{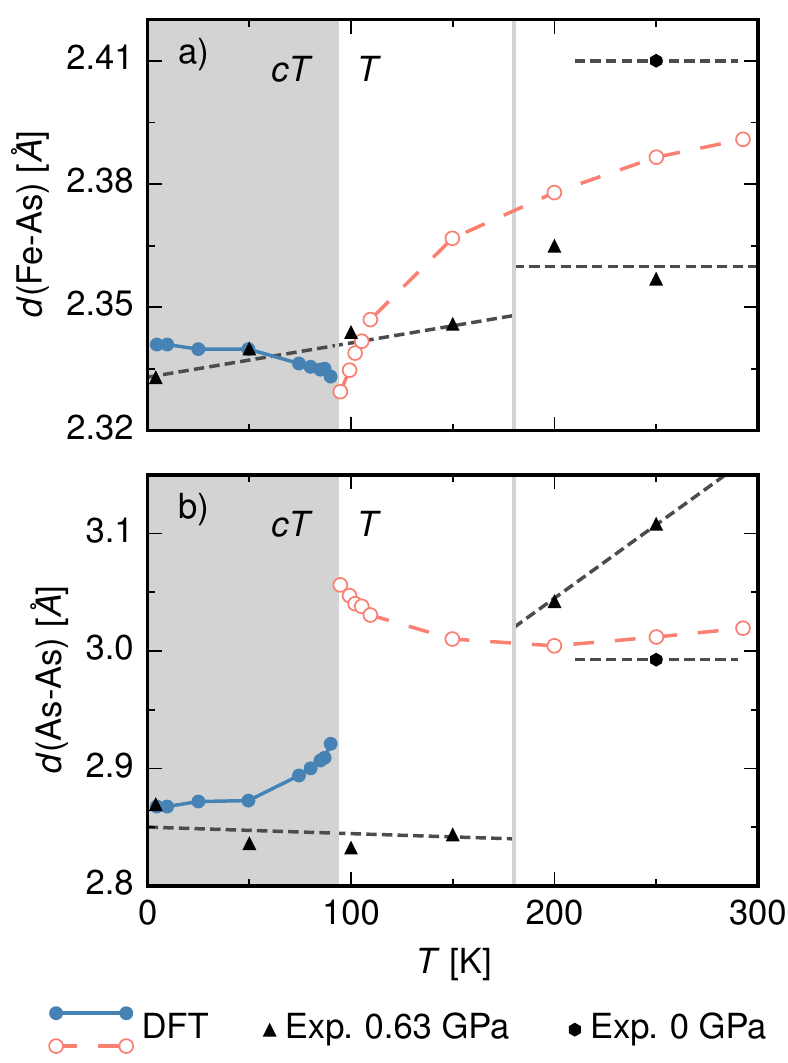}
\caption{Reconstruction of bond lengths in internally strained samples of 
Ca{\FeAs} from the $^{57}$Fe M{\"o}ssbauer 
spectroscopy. Shown are Fe-As bond lengths (a) and interlayer As-As bond lengths (b). 
The theoretical values were obtained within DFT (GGA).
Reprinted with permission from Ref.~\onlinecite{Budko2016}.} 
\label{fig:ca_bonds_mossbauer}
\end{figure}

$^{57}$Fe M{\"o}ssbauer spectroscopy of the samples, in conjunction with ab-initio calculations, 
provides additional insight into the microscopics of the collapsed tetragonal phase~\cite{Budko2016}. 
Since M{\"o}ssbauer spectroscopy probes properties intimately coupled to the electron charge density 
and the electric field gradients at the absorption nucleus~\cite{Greenwood1971} it provides valuable 
information about the immediate electronic environment of the iron nuclei. The Fe-As and interlayer As-As 
bond lengths deduced from M{\"o}ssbauer spectroscopy (see Fig.~\ref{fig:ca_bonds_mossbauer}) show a 
picture consistent with the behavior seen under pressure and provide insight into the physics behind 
the stabilization of the collapsed tetragonal phase. Namely, internally strained samples show relatively 
large interlayer As-As bond lengths, which further expand upon cooling. This results in charge 
saturation of the Fe-As bonds, which contract as the temperature is lowered. This process continues up to 
a point where it becomes energetically more favorable to transfer some of the charge from the Fe-As into 
the emptier interlayer As-As bond region, which prompts the formation of the interlayer As-As bonds and 
the formation of the collapsed tetragonal phase. 

\begin{figure}[tb]
\includegraphics[width=0.49\textwidth]{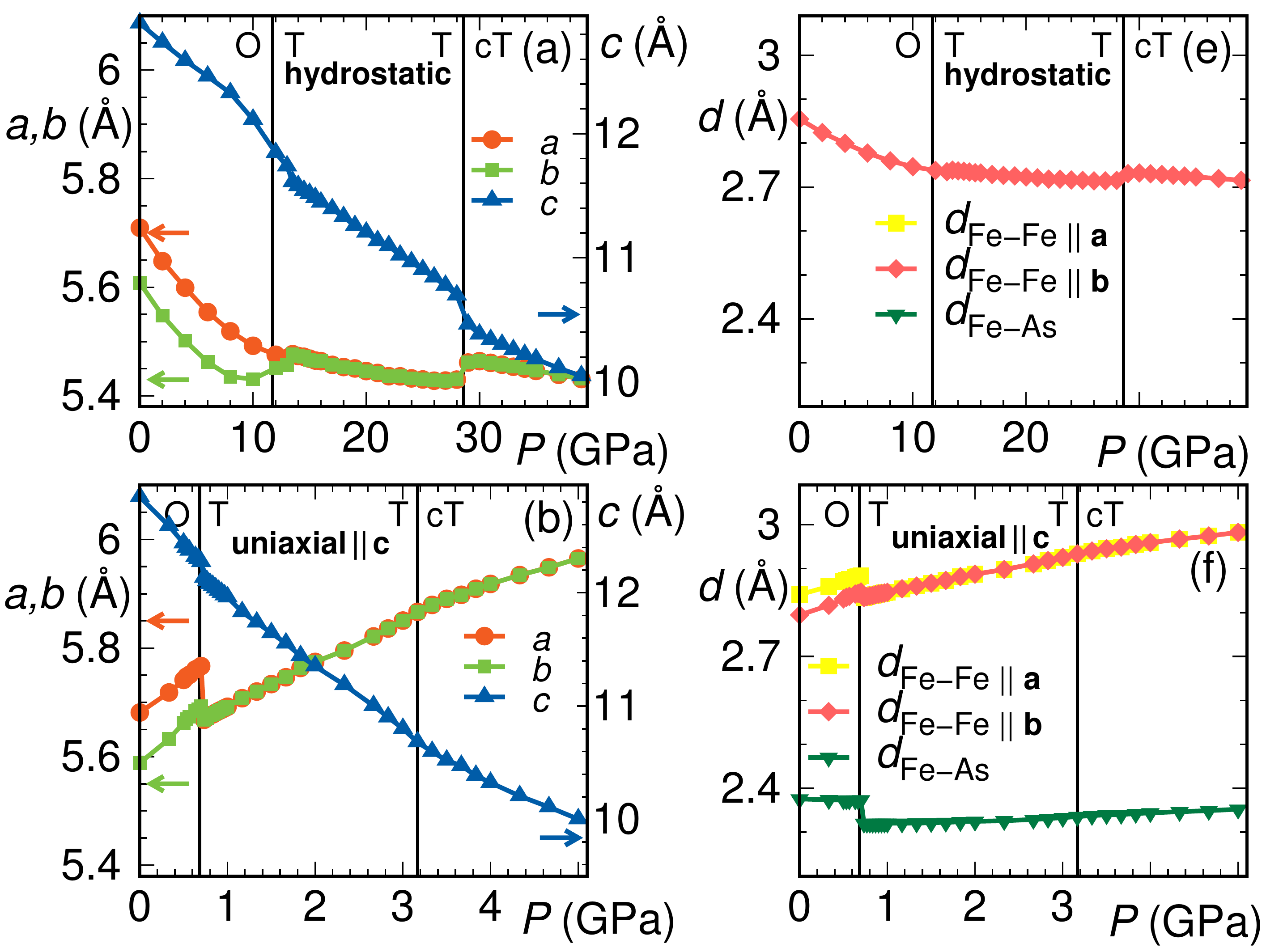}
\caption{Structure parameters of Ba{\FeAs} under application of hydrostatic (top 
row) and uniaxial pressure (bottom row). Shown are 
lattice parameters (a,b), and Fe-Fe and Fe-As bond lengths (c,d) calculated from
DFT (GGA). 
Reprinted with permission from Ref.~\onlinecite{Tomic2012}.}  \label{fig:ba_press_struct}
\end{figure}

In Ba{\FeAs} the rates of contraction of both the \caxis-axis and the 
\abplane-plane are faster with pressure than in Ca{\FeAs} [see 
Fig.~\ref{fig:ba_press_struct}(a)] as a consequence of a larger unit cell volume 
of Ba{\FeAs} due to the much larger size of the barium atom~\cite{Tomic2012}. 
Another consequence 
is that the formation of the collapsed tetragonal phase is delayed to much 
higher pressure of 28.6~GPa and is preempted by formation of an intermediate 
non-magnetic tetragonal phase at 11.75~GPa, which is consistent with other 
theoretical~\cite{Colonna2011,Zhang2009} and experimental~\cite{Kimber2009, 
Mittal2011} findings. 

In contrast to the sharp first order transition seen in Ca{\FeAs}, the 
transition in Ba{\FeAs} is much more gradual and almost second order. The 
examination of the electronic structure shows that, in the case of Ba{\FeAs}, 
hole Fermi surface pockets around the $\Gamma$ point get smaller and finally 
disappear, so that the Fermi surface is not nested any more. In consequence the 
spin density wave state is weakened beyond the point where magnetic order 
becomes unsustainable. Once the magnetic order disappears, Ba{\FeAs} becomes 
tetragonal again. 

\begin{figure}[tb]
\includegraphics[width=0.3\textwidth]{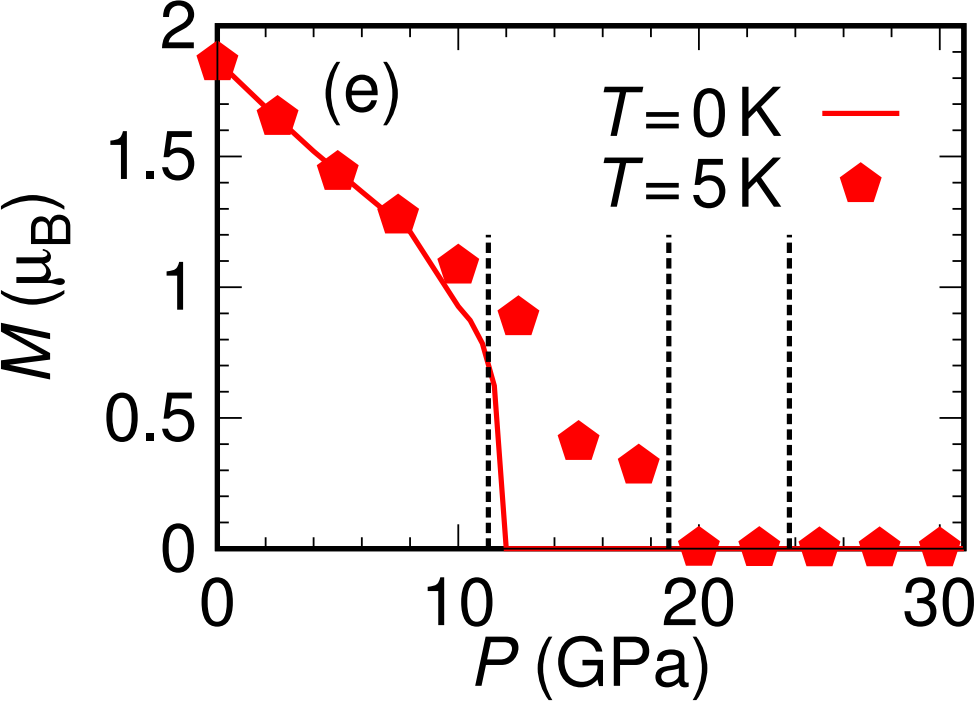}
\caption{Finite temperature magnetic moment of Ba{\FeAs} 
calculated from {\it ab-initio} molecular dynamics (DFT-GGA).
Reprinted with permission from Ref.~\onlinecite{Backes2013}.}
\label{fig:ba_mag_md}
\end{figure}

However, the interlayer As $p$-As $p$ bonds still do not form and the 
theoretical results suggest a scenario where some residual local magnetic 
moments remain, sustaining the magnetic fluctuations. The role of temperature in 
the magnetic fluctuations has been examined in more detail by finite temperature 
and pressure molecular dynamics~\cite{Backes2013}. In such finite temperature 
calculations both the crystal structure and the magnetic moments are allowed to 
fluctuate. The magnitude of magnetic moments at $T=5$~K is shown in 
Fig.~\ref{fig:ba_mag_md}. Around 12.5~GPa there is a transition to the low spin 
state, which persists up to about 20~GPa, where it is totally suppressed. This 
supports the fluctuating moment picture at finite temperature in the 
intermediate tetragonal phase. At $T=0$ and a pressure of 28.6~GPa the distance 
between the tri-layers is reduced enough so that the interlayer As-As bonds can 
form around the barium atom. 

The estimated ambient pressure bulk modulus is $67\pm 4$~GPa, which increases to 
$128\pm 3$~GPa for the intermediate tetragonal phase and up to $173\pm 2$~GPa 
for the collapsed tetragonal phase, in excellent agreement with the measured 
values~\cite{Mittal2011}.

We already mentioned the anisotropy of the 122 crystal structure and in 
particular the softness of the \caxis-axis. Therefore, we investigated uniaxial 
pressure effects along the \caxis-axis~\cite{Tomic2012}. The behavior of the 
unit cell of Ca{\FeAs} is shown in Fig.~\ref{fig:ca_press_struct}(b). There is a 
strong suppression of the \caxis-axis and homogeneous expansion in the 
\abplane-plane until 0.48~GPa, where the \caxis-axis collapses and the system 
enters the non-magnetic collapsed tetragonal phase. The order of magnitude 
reduction of transition pressure is in excellent agreement with the experimental 
observations~\cite{Prokes2010}. 

The electronic structure shows a complete suppression of the hole pockets around 
the $\Gamma$ point while Ca{\FeAs} is still in the orthorhombic magnetic phase. 
However, if we compare the Fe-As bond lengths [see 
Figs.~\ref{fig:ca_press_struct}(a) and (b)], we see that under the \caxis-axis 
uniaxial pressure the Fe-As bond suppression is slower, allowing for larger 
local moments due to the reduced crystal field splitting. Thus, although the 
Fermi surface is not nested, there is a large contribution of local moments 
maintaining the magnetic order.

For Ba{\FeAs}, the \caxis-axis uniaxial pressure also results in an order of 
magnitude reduction of the transition pressure, both for the intermediate 
tetragonal and collapsed tetragonal phase, from 11.6~GPa to 0.72~GPa and from 
28.6~GPa to 3.17~GPa, indicating that the larger size of Ba{\FeAs} does not have 
a detrimental effect for the anisotropy along the \caxis-axis. The Fe-As bond 
lengths are not suppressed below 2.3~$\AA$ with \caxis-axis uniaxial pressure, 
implying that the magnetic moments are more delocalized due to the more flat 
Fe-As tetrahedra. Increased sensitivity to the \caxis-axis uniaxial pressure is 
consistent with the experimentally observed behavior~\cite{Uhoya2010}.

Application of compressive strain along the \aaxis-axis obviously results in 
suppression of the orthorhombicity and magnetic moment both in Ca{\FeAs} and 
Ba{\FeAs}, but only up to a certain pressure, at which it becomes energetically 
more favorable to rotate the magnetic order by 90 degrees~\cite{Tomic2013}. At 
this point, the \aaxis~ and \baxis-axes switch their places and the 
orthorhombicity switches sign. This axis inversion is particularly interesting 
in the case of Ca{\FeAs}, where the magnetic moments have been shown to be quite 
fragile. It is also interesting to note that axis inversion in Ca{\FeAs} 
requires a larger pressure (0.67~GPa) than in Ba{\FeAs} (0.22~GPa). This is 
related to the fact that \caxis-axis uniaxial pressure is much more effective at 
magnetic moment suppression for Ca{\FeAs} than the uniaxial pressure applied in 
the \abplane-plane. This is the opposite of what is observed in Ba{\FeAs}. 
Similarly, tensile strain applied along the shorter \baxis-axis also results in 
axis inversion, with Ba{\FeAs} requiring lower pressure (-0.22~GPa) compared to 
Ca{\FeAs} (-0.33~GPa). This corresponds to the detwinning scenario where the 
orthorhombicity of one of the twin domains switches sign while the domain walls 
stay pinned. Our calculations showed a detwinning strain around 0.2~GPa, which 
is an order of magnitude larger than the detwinning strain needed in the 
tetragonal phase and is consistent with experimental observations~\cite{Chu2012, 
Blomberg2011}. Phenomenological Ginzburg-Landau modelling of the magneto-elastic 
physics leading to the axis inversion suggested that the larger reversal 
pressure in Ca{\FeAs} implies a larger magneto-elastic coupling in Ca{\FeAs} 
than in Ba{\FeAs}~\cite{Tomic2013}. This has a number of 
consequences~\cite{Fernandes2010, Cano2010, Barzykin2009}, one of which is that 
the magneto-structural transitions in Ca{\FeAs} are first-order, while in 
Ba{\FeAs} they are more second-order like. Application of tensile strain in the 
\abplane-plane~\cite{Tomic2013} results in suppression of hole pockets around 
the $\Gamma$ point and in more pronounced overall three-dimensionality of the 
Fermi surface. Indeed \caxis-axis uniaxial pressure also suppresses the $\Gamma$ 
point hole pockets due to the involved in-plane expansion [see 
Figs.~\ref{fig:ca_press_struct}(b) and \ref{fig:ba_press_struct}(b)].

\section{Effects of electronic correlations}
\label{sec:correlations}
In this section we focus our attention on the role of correlations in 
representative systems of the 111, 1111 and 122 families.

\begin{figure}[tb]
\includegraphics[width=1.0\linewidth]{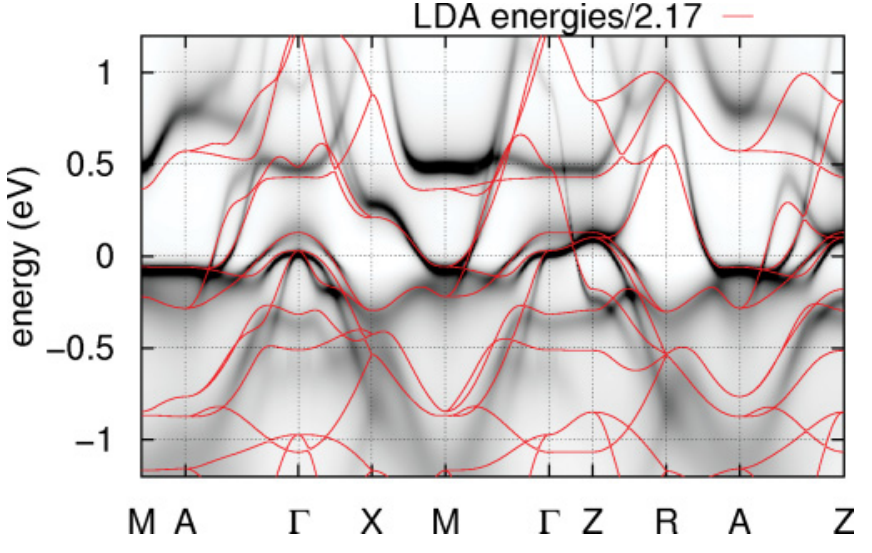}
\caption{Momentum resolved LDA+DMFT spectral function of LiFeAs compared
to the LDA dispersion (red lines). The LDA bands have been renormalized
by the orbitally averaged mass renormalization obtained from LDA+DMFT.
Reprinted with permission from Ref.~\onlinecite{Ferber2012}.}
\label{fig:LiFeAs:bs}
\end{figure}

{\it LiFeAs.-} We will first review the influence of electronic correlations in 
the iron-based superconductor LiFeAs and their effects on band structure and 
Fermi surface. Following our calculations based of LDA+DMFT~\cite{Ferber2012} in 
Fig.~\ref{fig:LiFeAs:bs} we compare the spectral function for LiFeAs as obtained 
from LDA+DMFT with its LDA counterpart at a temperature of $T=72.5$~K. The 
spectral function shows well defined excitations at the Fermi level, with 
increasing broadening due to the electronic correlations at higher binding 
energies, supporting the picture of well-defined quasiparticles at this 
temperature in this system. Therefore, this system shows characteristics of a 
Fermi-liquid state in a metal with moderate correlations without significant 
spectral weight transfer from the Fermi level to lower or higher binding 
energies.

\begin{figure}[tb]
\includegraphics[width=1.0\linewidth]{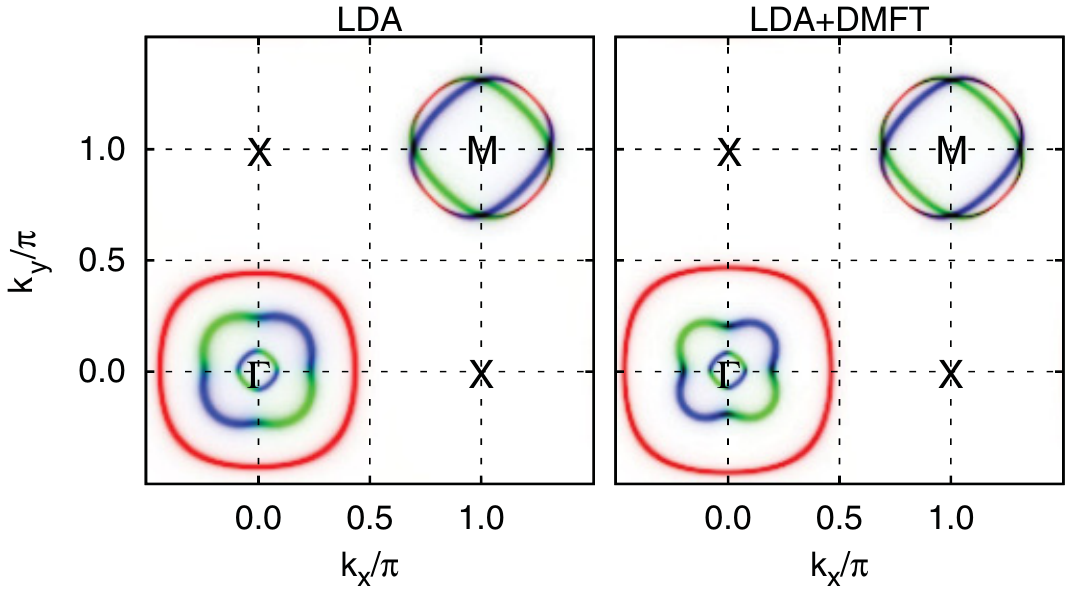}
\caption{LDA and LDA+DMFT Fermi surfaces at $k_z=0$ for LiFeAs. 
The color code labels the orbital
character: $d_{xy}$ (red), $d_{xz}$ (green) and $d_{xz}$ (blue).
Reprinted with permission from Ref.~\onlinecite{Ferber2012}.}
\label{fig:LiFeAs:fs}
\end{figure}

In Fig.~\ref{fig:LiFeAs:fs} we show the Fermi surface as obtained from LDA+DMFT, 
which shows hole pockets around the $\Gamma$-point and electron pockets around 
the $M$-point. The electronic correlations cause the shrinking of the middle 
$d_{xz/yz}$ hole pocket and an increase of the outer $d_{xy}$ pocket, whereas 
the shape of the electron pockets is hardly affected. This result indicates that 
the electronic correlations tend to weaken Fermi surface nesting in this 
material or even might suppress it. 

\begin{figure}[tb]
\includegraphics[width=1.0\linewidth]{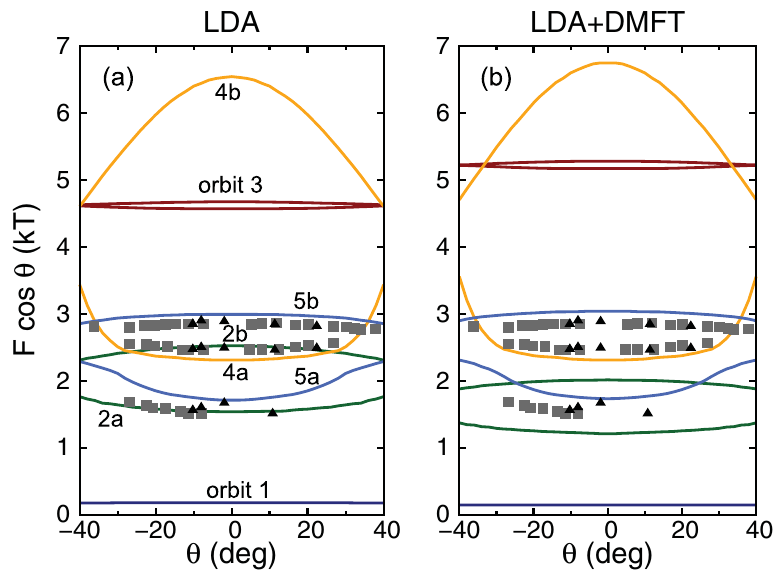}
\caption{dHvA frequencies for LiFeAs as a function of magnetic field angle.
The solid lines refer to the theoretical calculation, while the points refer 
to the experimental data from Ref.~\onlinecite{Putzke2012}.
The theoretical data was obtained by calculating the angle-dependent extremal 
cross-sections of the Fermi surface cylinders from the DFT (left) or LDA+DMFT 
(right) calculation. Reprinted with permission from 
Ref.~\onlinecite{Ferber2012}.}
\label{fig:LiFeAs:dhva}
\end{figure}

For a more quantitative comparison we calculated the dHvA frequencies from 
LDA+DMFT. The dHvA frequencies correspond to the extremal size of the Fermi 
surface pockets at a given angle with respect to the $k_z$ axis. In 
Fig.~\ref{fig:LiFeAs:dhva} we compare our theoretical results to the 
experimental measurements from Ref.~\onlinecite{Putzke2012}. Despite a few small 
differences, LDA seems to  agree with experiment quite well. 

Inclusion of 
electronic correlations induces a shrinking of the middle hole pocket as a 
downward shift of the corresponding frequency response, and an upward shift of 
the enlarged outer pocket. Ref.~\onlinecite{Putzke2012} assigns the measured 
frequencies to the electron Fermi surface sheets, where the two higher 
frequencies are assigned to orbits 5b and 4a, and the lowest frequency is 
assumed to originate from orbit 5a~\cite{ColdeaPrivate}. 

Our results support 
this interpretation: while the orbits 2a/5a and 2b/4a are of similar size in the 
LDA calculation, the hole pockets are modified and the near degeneracy in the de 
Haas-van Alphen frequency plot is lifted. Therefore, the electron orbits 2a and 
2b are unlikely to give rise to the measured frequencies, as their sizes are 
rather different from the measured data. This finding reconciles theory and 
experiment. The shrunk middle hole pocket is only seen in ARPES, which finds a 
correlated metal with poor nesting together with sizable mass renormalization. 
In contrast, the dHvA measurement resolves the (lighter) electron pocket sizes 
in LiFeAs that almost do not change under inclusion of correlation. Recent 
LDA+DMFT calculations for LiFeAs~\cite{Yin2011,Geunsik2012} show the same trends 
as our results~\cite{Ferber2012}.

%%%%%%%%%%%%%%%%%%%%%%%%%%%%%%%%%%%%%%%%%%%%%%%%%%%%%%%%%%%%%%%%%%%%%%%%%%%%%
% LaFePO and LiFeP discussion %%%%%%%%%%%%%%%%%%%%%%%%%%%%%%%%%%%%%%%%%%%%%%%
%%%%%%%%%%%%%%%%%%%%%%%%%%%%%%%%%%%%%%%%%%%%%%%%%%%%%%%%%%%%%%%%%%%%%%%%%%%%%

\begin{figure}[tb]
\includegraphics[width=1.0\linewidth]{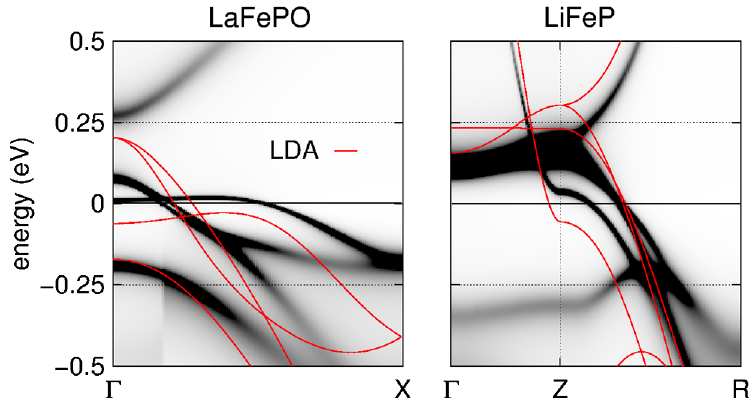}
\caption{Momentum resolved LDA+DMFT spectral function of LaFePO (left) 
and LiFeP (right)
together with the LDA bands close to the Fermi surface topology change.
Reprinted with permission from Ref.~\onlinecite{Ferber2012a}.}
\label{fig:LaFePO:bs}
\end{figure}

{\it LiFeP and LaFePO.-} As a comparison study, we  review the LDA+DMFT 
calculations for LaFePO and LiFeP~\cite{Ferber2012a}. Fig.~\ref{fig:LaFePO:bs} shows 
the momentum resolved spectral function for both materials and the comparison to 
the LDA result. Both compounds show a distinctive change in the topology of the 
Fermi surface with a  hole pocket of Fe $dz_2$ orbital character changing from a 
closed shape in LDA to an open shape in LDA+DMFT. This change of topology occurs 
around the $\Gamma$ (Z) point in LaFePO (LiFeP). 

\begin{figure}[tb]
\includegraphics[width=1.0\linewidth]{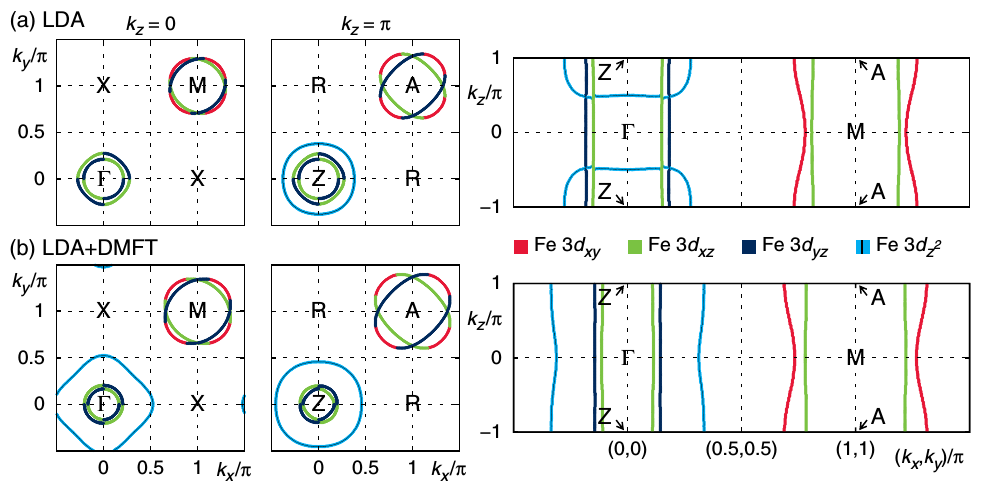}
\caption{Fermi surfaces for LaFePO in the $k_z=0,\pi$ plane (left) and 
$k_x=k_y$ plane (right). The upper row shows the result obtained from
DFT, while the lower row shows the modified Fermi surface including electronic 
correlations on the LDA+DMFT level. The colors indicate the orbital character.
Reprinted with permission from Ref.~\onlinecite{Ferber2012a}.}
\label{fig:LaFePO:fs}
\end{figure}

This effect is clearly visible in the calculated Fermi surface. 
Fig.~\ref{fig:LaFePO:fs} shows the appearance of an additional outer hole pocket 
at $\Gamma$ in LaFePO and an inner hole pocket at Z in LiFeP~\cite{Ferber2012a}. 
As discussed by Kemper {\it et al.}~\cite{Kemper2010}, this might promote a 
nodal gap and weaken the pairing strength, in turn also lowering the 
superconducting transition temperature. In conclusion, while LiFeP and LaFePO 
are less correlated than LiFeAs (band renormalizations and effective masses are 
smaller), correlation effects are crucial to understand the topology of the 
Fermi surface.

%%%%%%%%%%%%%%%%%%%%%%%%%%%%%%%%%%%%%%%%%%%%%%%%%%%%%%%%%%%%%%%%%%%%%%%%%%%%%
% CaFe2As2 discussion %%%%%%%%%%%%%%%%%%%%%%%%%%%%%%%%%%%%%%%%%%%%%%%%%%%%%%%
%%%%%%%%%%%%%%%%%%%%%%%%%%%%%%%%%%%%%%%%%%%%%%%%%%%%%%%%%%%%%%%%%%%%%%%%%%%%%
\begin{figure}[tb]
\includegraphics[width=1.0\linewidth]{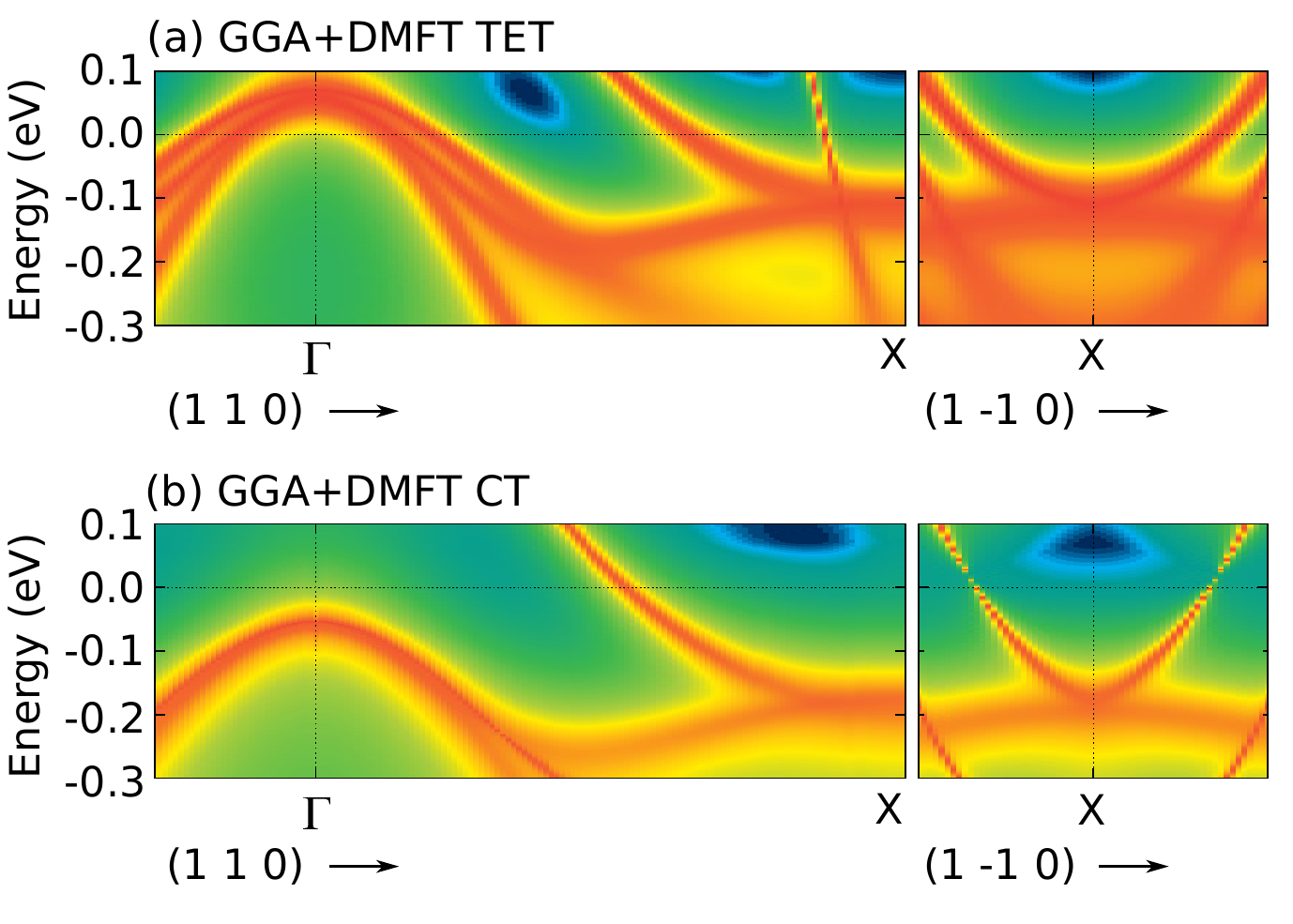}
\caption{Momentum resolved LDA+DMFT spectral function for Ca{\FeAs} in the tetragonal
and collapsed tetragonal phase.
Reprinted with permission from Ref.~\onlinecite{Diehl2014}.}
\label{fig:CaFe2As2:bs}
\end{figure}

{\it CaFe$_2$As$_2$.-} We proceed now with the 122 family and the interplay 
between correlation effects and pressure. We consider here Ca{\FeAs} as a 
representative system, where we review the effects of correlation between the 
tetragonal and the collapsed tetragonal phase as discussed in 
Refs.~\onlinecite{Liu2009B, Soh2013}. In contrast to the previous materials, the 
topology of the Fermi surface in both phases remains nearly unaffected upon the 
inclusion of electronic correlations.
This manifests in the orbital-selective mass renormalizations of a factor of 
$1.3$ to $1.7$ introduced by LDA+DMFT. 

Nonetheless, these results improve the agreement of the calculations with ARPES 
experiments~\cite{Dhaka2014} compared to the pure LDA result, since the 
bandwidth renormalization obtained in experiment is quite well reproduced by the 
calculation. In Fig.~\ref{fig:CaFe2As2:bs} we show the results for the 
momentum-resolved spectral function in both the tetragonal and collapsed 
tetragonal phase. The tetragonal phase shows three hole pockets at the 
$\Gamma$-point ant two electron pockets at the $M$-point, whereas in the 
collapsed tetragonal phase the hole pockets are pushed below the Fermi level and 
thus vanish, while the electron pockets increase significantly in size. 

The influence of the tetragonal to collapsed tetragonal transition on the 
orbital-dependent effective masses manifests in a decrease of electronic 
correlation effects, where the Fe $3d_{xy}$ orbital changes from being the most 
strongly correlated orbital in the tetragonal phase to being the least 
correlated one in the collapsed tetragonal phase. This change is due to the 
increase in hybridization of the Fe $3d$ orbitals after the structural collapse. 
The decreased distance of the Fe-As layers enforces hybridization of the Fe 
3$d_{xy}$-Fe 3$d_{xy}$ as well as Fe 3$d_{xy}$-As 4$p_x$ and 4$p_y$ orbitals. 
This reduces the localization of the electrons and renders the Fe 3$d_{xy}$ less 
localized (and thus less correlated). In conclusion, electronic correlation 
effects are important even for weakly correlated pnictides like Ca{\FeAs} in 
order to understand the orbital-selective mass renormalizations that are seen in 
ARPES measurements.

%%%%%%%%%%%%%%%%%%%%%%%%%%%%%%%%%%%%%%%%%%%%%%%%%%%%%%%%%%%%%%%%%%%%%%%%%%%%%
% KFe2As2 discussion %%%%%%%%%%%%%%%%%%%%%%%%%%%%%%%%%%%%%%%%%%%%%%%%%%%%%%%%
%%%%%%%%%%%%%%%%%%%%%%%%%%%%%%%%%%%%%%%%%%%%%%%%%%%%%%%%%%%%%%%%%%%%%%%%%%%%%
{\it KFe$_2$As$_2$.-} We now move to the description of our results obtained for 
the hole-doped iron-pnictide superconductor K{\FeAs}~\cite{Backes2014}. 
Especially for this material, DFT calculations are insufficient to 
satisfactorily describe angle-resolved photoemission (ARPES) measurements as 
well as observed de Haas van Alphen (dHvA) frequencies. 

In our LDA+DMFT calculations reported in Ref.~\onlinecite{Backes2014} we find 
that K{\FeAs} is a moderately to strongly correlated metal with a mass 
renormalization factor of the Fe $3d$ orbitals between 1.6 and 2.7. The observed 
strong flattening of electronic bands due to the renormalization is a possible 
explanation for the spread of experimental results in this compound in terms of 
extreme sensitivity to the experimental stoichiometry. We find significant 
modifications in the size and shape of the Fermi surface pockets, which in this 
system are only composed of hole pockets. 

\begin{figure}[tb]
\includegraphics[width=0.475\linewidth]{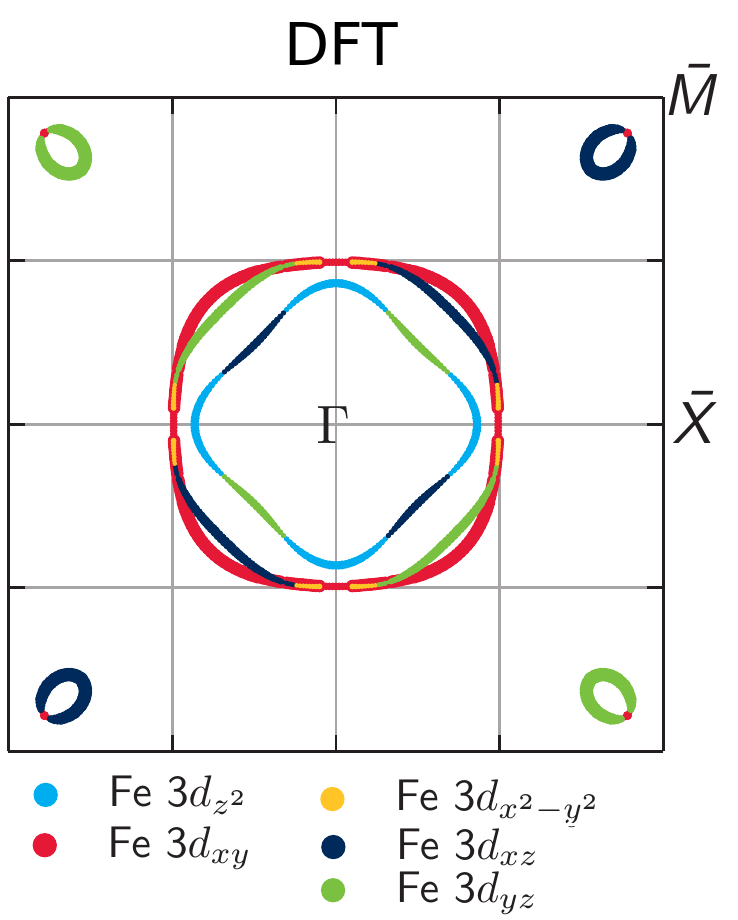}
\includegraphics[width=0.48\linewidth]{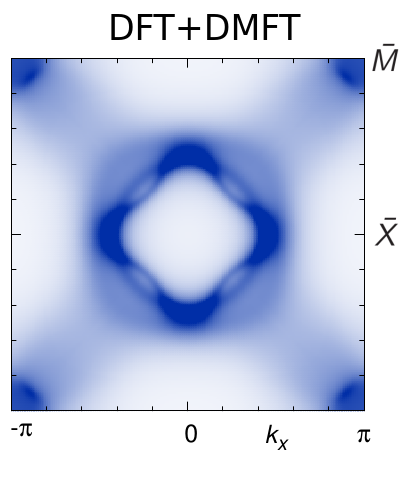}
\caption{Fermi surface at $k_z=0$ of K{\FeAs} as obtained from DFT (left) and
the momentum resolved spectral function at the Fermi level as obtained
from LDA+DMFT (right).
Reprinted with permission from Ref.~\onlinecite{Backes2014}.}
\label{fig:KFe2As2:fs}
\end{figure}

In Fig.~\ref{fig:KFe2As2:fs} we show a comparison between the Fermi surface 
obtained from DFT and LDA+DMFT at $k_z=0$. The DFT result contains a middle hole 
cylinder at $\Gamma$, which is too large compared to 
ARPES~\cite{Yoshida2011,Yoshida2014}. Correlations within LDA+DMFT increase the 
size of the outer cylinder while reducing the size of the middle, and also 
modifys the shape of the inner cylinder, which greatly improves the agreement 
with the ARPES measurements. Additionally, we predict a topological change with 
respect to DFT calculations, namely the opening of an inner hole cylinder at the 
$Z$ point. As a result, we also found that our calculated dHvA frequencies are 
modified by electronic correlations and qualitatively agree with experimental 
data~\cite{Terashima2010,Terashima2013A}. Furthermore, the intersection nodes on 
the inner two hole cylinders offer a natural explanation for magnetic breakdown 
orbits observed in the dHvA measurements~\cite{Terashima2013A}.

On this basis, we argue that correlation effects are important and a necessary 
ingredient in understanding the electronic structure in K{\FeAs}, as well as the 
presently under debate nature of the superconducting state in K{\FeAs}.

%%%%%%%%%%%%%%%%%%%%%%%%%%%%%%%%%%%%%%%%%%%%%%%%%%%%%%%%%%%%%%%%%%%%%%%%%%%%%
% hole-doped 122 discussion %%%%%%%%%%%%%%%%%%%%%%%%%%%%%%%%%%%%%%%%%%%%%%%%%
%%%%%%%%%%%%%%%%%%%%%%%%%%%%%%%%%%%%%%%%%%%%%%%%%%%%%%%%%%%%%%%%%%%%%%%%%%%%%
{\it $A$Fe$_2$As$_2$ ($A={\rm K}$, Rb, Cs).-} In our previous 
studies, for example in Ca{\FeAs}~\cite{Diehl2014} and 
K{\FeAs}~\cite{Backes2014}, which showed that: (i) a compression of the unit 
cell in general reduces the strength of electronic correlations and (ii) the 
hole doping of the parent compound Ba{\FeAs} by substituting Ba by K yields a 
moderately to strongly correlated system with notable correlation effects in the 
electronic structure. 

This naturally raises the question whether a decompression, i.e. increase of the 
interatomic distances in the unit cell increases the electronic correlations as 
opposed to a reduction found under a compression of the unit cell. Guided by 
this question we investigate via LDA+DMFT the manifestation of correlation 
effects in a wide range of binding energies in the hole-doped family of 
Fe-pnictides $A$Fe$_2$As$_2$ ($A={\rm K}$, Rb, Cs) as well as the fictitious 
Fr{\FeAs} and $a$-axis stretched Cs{\FeAs}. This choice of systems allows for a 
systematic analysis of the interplay of Hund's coupling $J_H$ and on-site 
Coulomb repulsion $U$ in multi-orbital Fe-pnictides under {\it negative} 
pressure, described in detail in  Ref.~\onlinecite{Backes2015}.

\begin{figure}[tb]
\includegraphics[width=\linewidth]{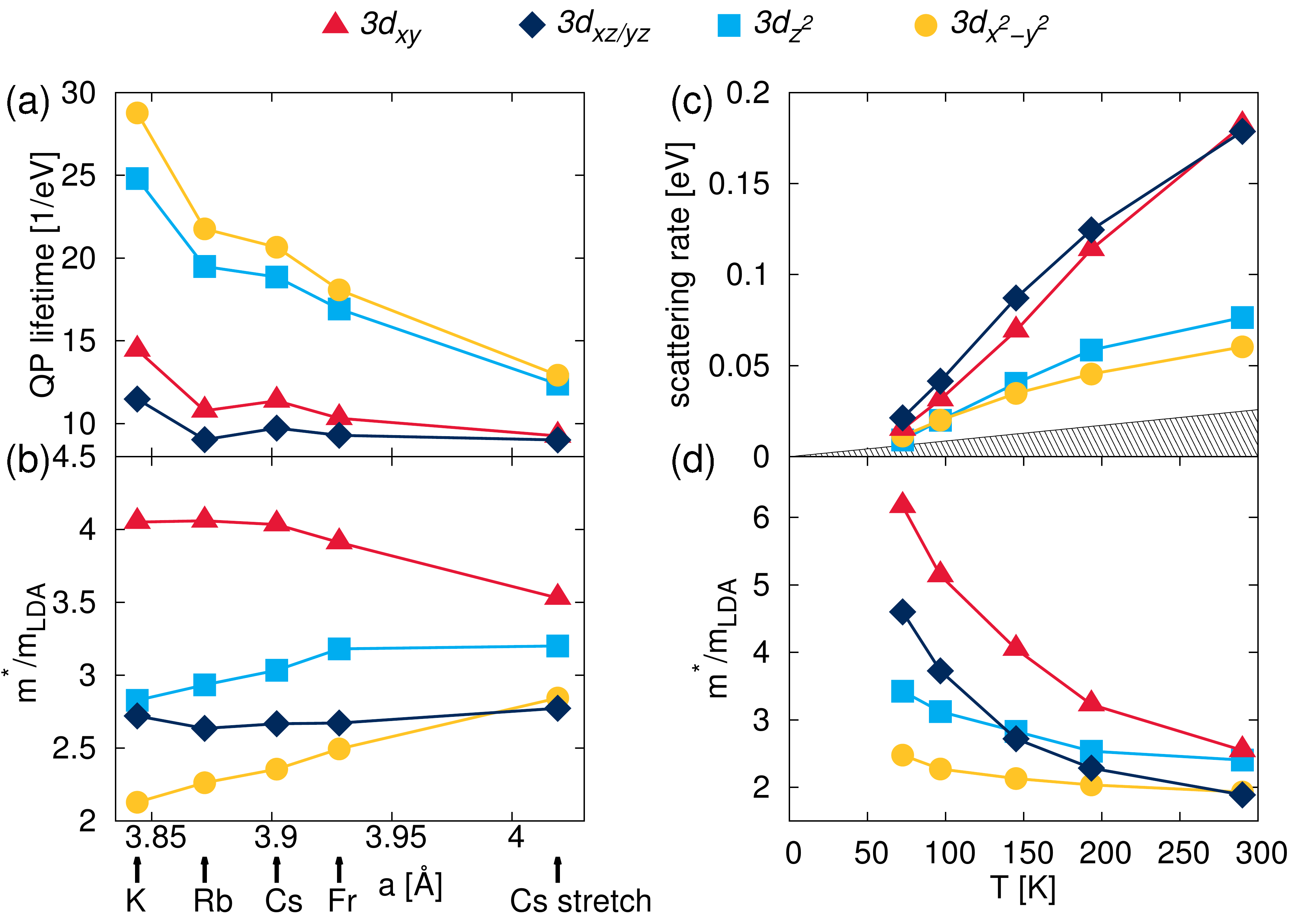}
\caption{The quasiparticle lifetime (a) and mass enhancements (b) as obtained 
from LDA+DMFT as a function of increasing atomic  radius in the $A${\FeAs} 
($A$=K,Rb,Cs,Fr). The temperature dependence of the scattering rate (c) and mass 
enhancement (d) for the example of K{\FeAs} shows that these systems are quite 
deep in the incoherent regime with a coherence temperature of about $50$~K. 
Reprinted with permission from Ref.~\onlinecite{Backes2015}.}
\label{fig:122series:meff_life}
\end{figure}

When increasing the ionic size of the alkali metal,
we observe (i) a non-trivial change 
in the iron $3d$ hoppings, 
(ii) an increase of orbitally-selective correlations and 
(iii) transfer of incoherent spectral weight to high-binding energies.
 We do not find 
the typical lower Hubbard-band, but rather characteristic features of a 
Hund's metal. This is especially prominent in $a$-stretched Cs{\FeAs}. We also 
find that the coherent/incoherent electronic behavior of the systems depends, apart 
from temperature, also strongly on $J_H$. To elucidate this, we show the 
quasiparticle lifetime and the orbitally resolved effective masses as a function 
of atomic radius in Fig.~\ref{fig:122series:meff_life}(a) and (b). 

We find a strong 
suppression of the quasiparticle lifetimes for larger atomic radius, as well as 
an overall increase in electronic correlations. From our analysis of the 
electronic properties in a wide range of binding energies, we come to the 
conclusion that along the isoelectronic doping series $A$Fe$_2$As$_2$ ($A={\rm 
K}$, Rb, Cs), and also for the fictitious Fr{\FeAs} and $a$-axis stretched Cs{\FeAs}, 
correlation and incoherence of the Fe $3d$ orbitals increase, albeit orbitally 
selective. 

These systems show distinctive features of Hund's metals, i.e. the 
Hund's coupling $J_H$ plays a major role in the strength of correlations and 
especially coherence. Therefore, these materials are 
much more incoherent than expected from the value of the Coulomb repulsion $U$ 
alone. While the most correlated orbitals ($d_{xy}$) show features that resemble 
those of being close to an orbital selective Mott transition, especially for 
$a$-stretched Cs{\FeAs}, the system is quite deep in the incoherent bad metal 
regime with a finite spectral weight at the Fermi level even when we vary the 
interaction parameters in the range from $U=4$~eV, $J_H=0.8$~eV and $U=6$~eV, 
$J_H=1.2$~eV. This actually shows that the systems are not close to an orbital 
selective Mott transition, but rather are highly incoherent due to the 
suppression of orbital fluctuations by the Hund's coupling at the 
temperature considered. 

We 
predict that by increasing the Fe-Fe distance experimentally in the most correlated and 
incoherent system Cs{\FeAs}, e.g. by stretching, will induce an orbital dependent 
increase in correlations and incoherence of the Fe $3d$ orbitals, where the Fe 
$3d_{z^2}$ and Fe $3d_{xy}$ orbitals are strongly but not fully localized and 
the other Fe $3d$ orbitals retain a bad metallic behavior. From our results of 
the temperature dependence of the scattering rate shown in 
Fig.~\ref{fig:122series:meff_life}(c) we estimate the coherence temperature  to be  
located around $50$~K in K{\FeAs} and even lower for Rb{\FeAs} and Cs{\FeAs}. This agrees 
qualitatively with experimental 
observations~\cite{Hardy2013,Boehmer_thesis2014}. 

The incoherent properties also 
render the usual way of obtaining the mass enhancements by the slope of the 
self-energy invalid, which assumes Fermi liquid properties. These are clearly violated if the imaginary 
part of the self-energy takes on a finite value for $\omega\rightarrow 0$, i.e. 
has a significant scattering rate.

Therefore, we conclude that especially the 
hole doped end systems of the 122 iron pnictide family K{\FeAs}, Rb{\FeAs} and 
Cs{\FeAs}, as well as the $a$-axis stretched Cs{\FeAs} are a valuable test bed to 
study the features of strongly correlated Hund's metals and orbital-selective 
bad metallicity and its interplay with superconductivity.

%%%%%%%%%%%%%%%%%%%%%%%%%%%%%%%%%%%%%%%%%%%%%%%%%%%%%%%%%%%%%%%%%%%%%%%%%%%%%
%%%%%%%%%%%%%%%%%%%%%%%%%%%%%%%%%%%%%%%%%%%%%%%%%%%%%%%%%%%%%%%%%%%%%%%%%%%%%
% End of correlation sections %%%%%%%%%%%%%%%%%%%%%%%%%%%%%%%%%%%%%%%%%%%%%%%
%%%%%%%%%%%%%%%%%%%%%%%%%%%%%%%%%%%%%%%%%%%%%%%%%%%%%%%%%%%%%%%%%%%%%%%%%%%%%
%%%%%%%%%%%%%%%%%%%%%%%%%%%%%%%%%%%%%%%%%%%%%%%%%%%%%%%%%%%%%%%%%%%%%%%%%%%%%

%========================================================================================================
%  Superconductivity
%========================================================================================================

\section{Superconductivity}
\label{sec:superconductivity}
In this section we concentrate on our work on superconductivity in extremely 
hole- and electron-doped iron pnictides and chalcogenides. Our investigations 
were driven by experimental progress in application of pressure and sample 
preparation.  
In the following subsections we 
review the experimental situation in extremely hole-doped iron pnictides under 
high pressures and the current status of iron selenide and intercalates. We put our 
work into the context provided by the relevant experiments.

%
% EXTREMELY HOLE DOPED IRON PNICTIDES UNDER PRESSURE
%
{\it Extremely hole-doped iron pnictides under pressure.-}
The phase diagram, and in particular the normal state properties, of the 
$A$\FeAs ($A=$~Ca,~Ba,~K,~Rb,~Cs,~Fr) family of materials have been discussed 
extensively in the previous sections. In this subsection we concentrate on the 
K\FeAs~material, which superconducts below a critical temperature $T_c = 
3.4~\mathrm{K}$~\cite{Tafti2013}. For moderate pressures a V-shaped dependence 
of $T_c$ has been observed in some experiments~\cite{Tafti2013, Tafti2014, 
Tafti2015}, while no such behavior is found under slightly different 
experimental conditions~\cite{Ying2015, Nakajima2015}. Surprisingly, at high 
pressures a superconducting phase with $T_c$ of up to $12~\mathrm{K}$ was 
found~\cite{Ying2015, Nakajima2015}, 
which is linked to a structural collapse as it is known from Ca\FeAs~and 
Ba\FeAs.

From our previous work we know that correlation effects are extremely 
important in K\FeAs~and that even the Fermi surface from pure DFT is 
incorrect~\cite{Backes2014}. At large pressures one can however expect the 
electronic bandwidth to increase and correlation effects to diminish in 
consequence. We find that this is indeed the case for the high pressure 
collapsed phase of K\FeAs~based on lattice parameters taken from 
Ref.~\onlinecite{Ying2015}. In fact, the Fermi surface of the collapsed phase is 
insensitive to inclusion of correlation effects~\cite{Guterding2015}. 

\begin{figure}[tb]
\includegraphics[width=\linewidth]{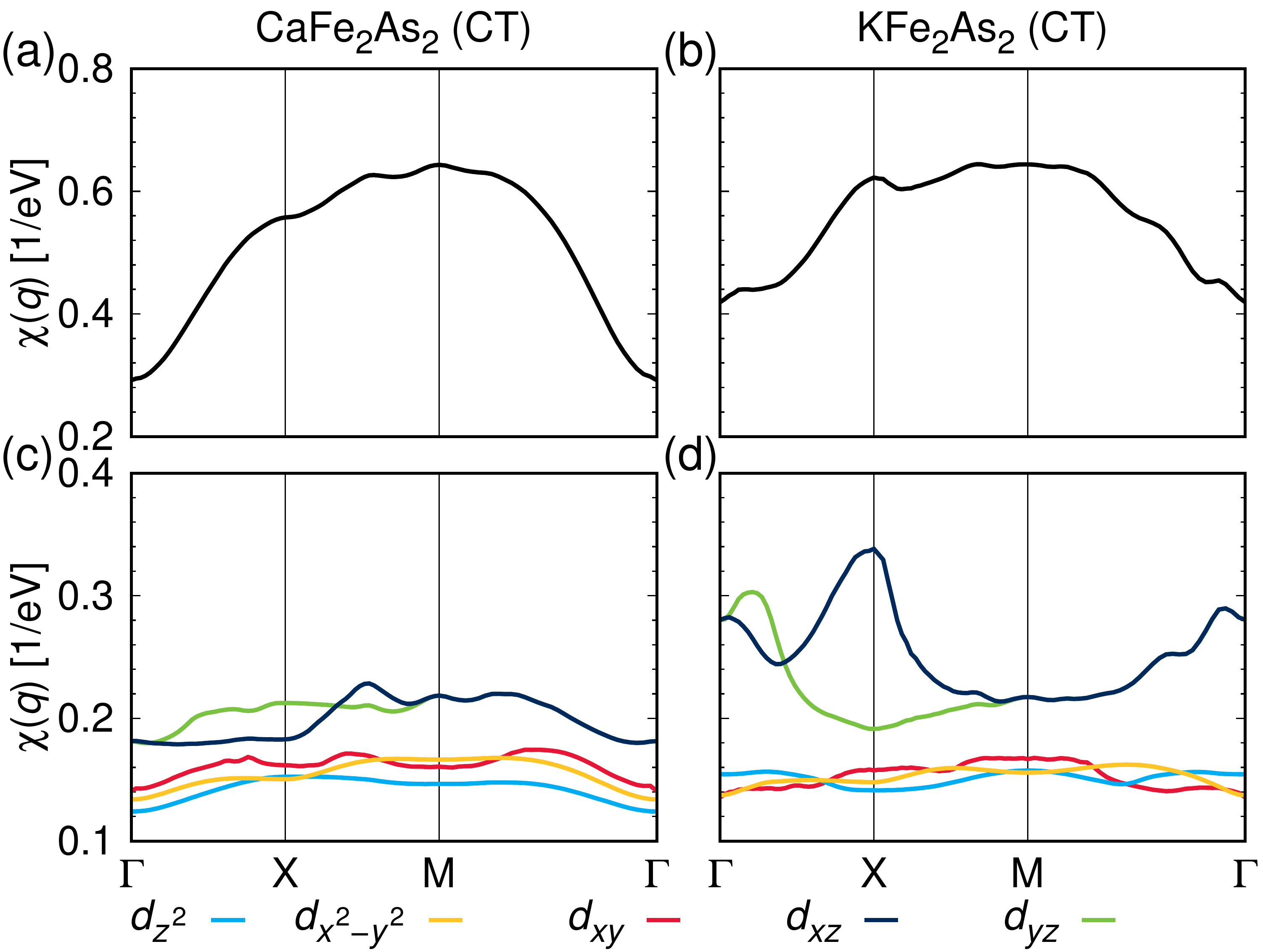}
\caption{Summed static susceptibility (top) and its
  diagonal components $\chi_{aa}^{aa}$ (bottom) in the eight-band
  tight-binding model for [(a) and (c)] CaFe$_2$As$_2$ and [(b) and
  (d)] KFe$_2$As$_2$ in the one-Fe Brillouin zone, based on DFT (GGA).
  The colors identify the Fe 3$d$ states.
Reprinted with permission from Ref.~\onlinecite{Guterding2015}.}
\label{fig:KFe2As2:susceptibilitycollapsed}
\end{figure}

\begin{figure}[tb]
\includegraphics[width=\linewidth]{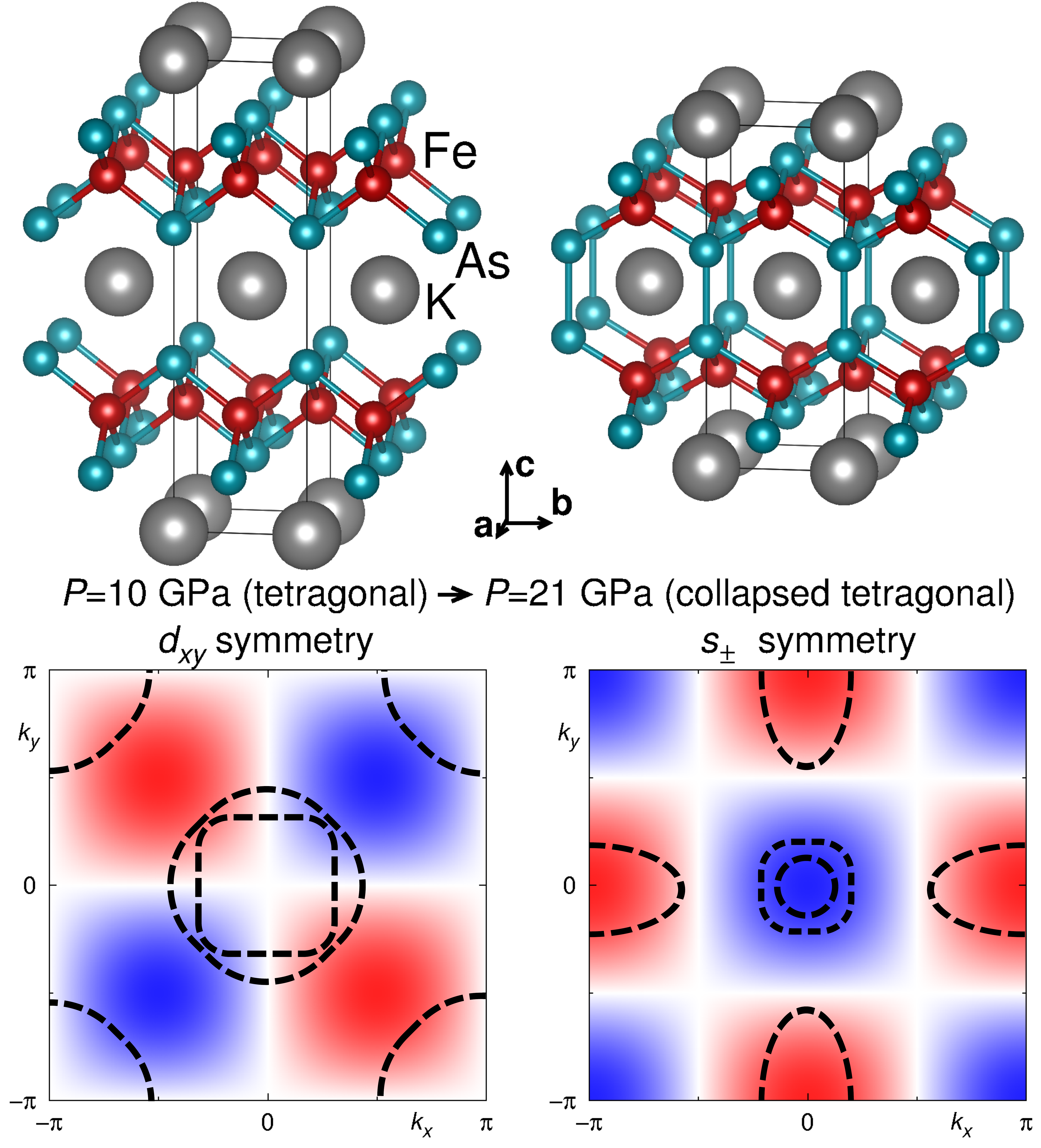}
\caption{Crystal structure, schematic Fermi surface (dashed lines) and schematic 
superconducting gap function (background color) of KFe$_2$As$_2$ in the one-Fe 
Brillouin zone before and after the volume collapse. The Lifshitz transition
associated with the formation of As 
4$p_z$-As 4$p_z$ bonds in the CT phase changes the superconducting pairing 
symmetry from $d_{xy}$ to $s_\pm$. Reprinted with permission from 
Ref.~\onlinecite{Guterding2015}.}
\label{fig:KFe2As2:orderparameters}
\end{figure}

Performing DFT 
 calculations in the GGA approximation
we furthermore find that a Lifshitz transition is 
associated with the structural collapse~\cite{Guterding2015}. The electronic 
structure on the low-pressure side of the phase transition is qualitatively the 
same as that at zero pressure,
 while in the collapsed phase it is similar to that of 
Ca\FeAs, but with additional small hole pockets in the Brillouin zone center. We 
showed that this difference is crucial for establishing nesting with wave vector 
$X = (\pi, 0)$, as evidenced in the static spin-susceptibility (see 
Fig.~\ref{fig:KFe2As2:susceptibilitycollapsed}). From RPA we have predicted
unconventional superconductivity with a sign-changing $s_\pm$ order parameter to 
be realized in the collapsed phase of K\FeAs~\cite{Guterding2015}, while the
non-collapsed phase is a $d$-wave superconductor (see also 
Refs.~\onlinecite{Thomale2011A, Maiti2011} for ambient pressure results). Therefore, 
the structural collapse does not only induce a Lifshitz transition, but also 
changes the symmetry of the superconducting state from $d$-wave to extended 
$s$-wave (see Fig.~\ref{fig:KFe2As2:orderparameters}).

The observation that small hole pockets at the Brillouin zone center re-emerge 
under pressure could also explain why superconductivity has been found in the 
collapsed phase of Ca\FeAs~under uniaxial pressure, but not under hydrostatic 
conditions. We have shown previously that uniaxial pressure quickly leads to the 
re-emergence of hole pockets at the Brillouin zone center in this 
compound~\cite{Tomic2012}.

%
% PRESSURE AND DOPING IN IRON SELENIDE
%

{\it Pressure and doping in iron selenide.-}
Iron-based superconductors are not only amenable to modification by application 
of pressure, but also by charge doping. The effects of both are evidenced 
prominently in the iron selenide material. 

While unpressurized bulk FeSe has a $T_c$ of 8-10~K, pressure enhances the 
critical temperature to about $T_c \sim 40~\mathrm{K}$~\cite{Medvedev2009}. This 
comes entirely unexpected, as most other iron-based superconductors are thought 
to have an antiferromagnetic parent state, which can be suppressed by 
application of pressure. Pressing further, the superconducting critical 
temperature also decreases in these compounds. In FeSe instead of an 
antiferromagnetic dome, a large $C_2$-symmetric nematic region is found, where 
the crystal structure is already orthorhombic, but no static magnetism is 
realized.

We recently found that FeSe is subject to an unexpected magnetic frustration not 
found in other iron-based superconductors. In Ref.~\onlinecite{Glasbrenner2015}
we show that a $J_1 - J_2 - J_3$ 
exchange model with additional biquadratic term $K$ accurately describes the 
non-monotonic pressure dependence of superconductivity and the  orbital 
ordered nematic region in FeSe.

\begin{figure}[tb]
\includegraphics[width=\linewidth]{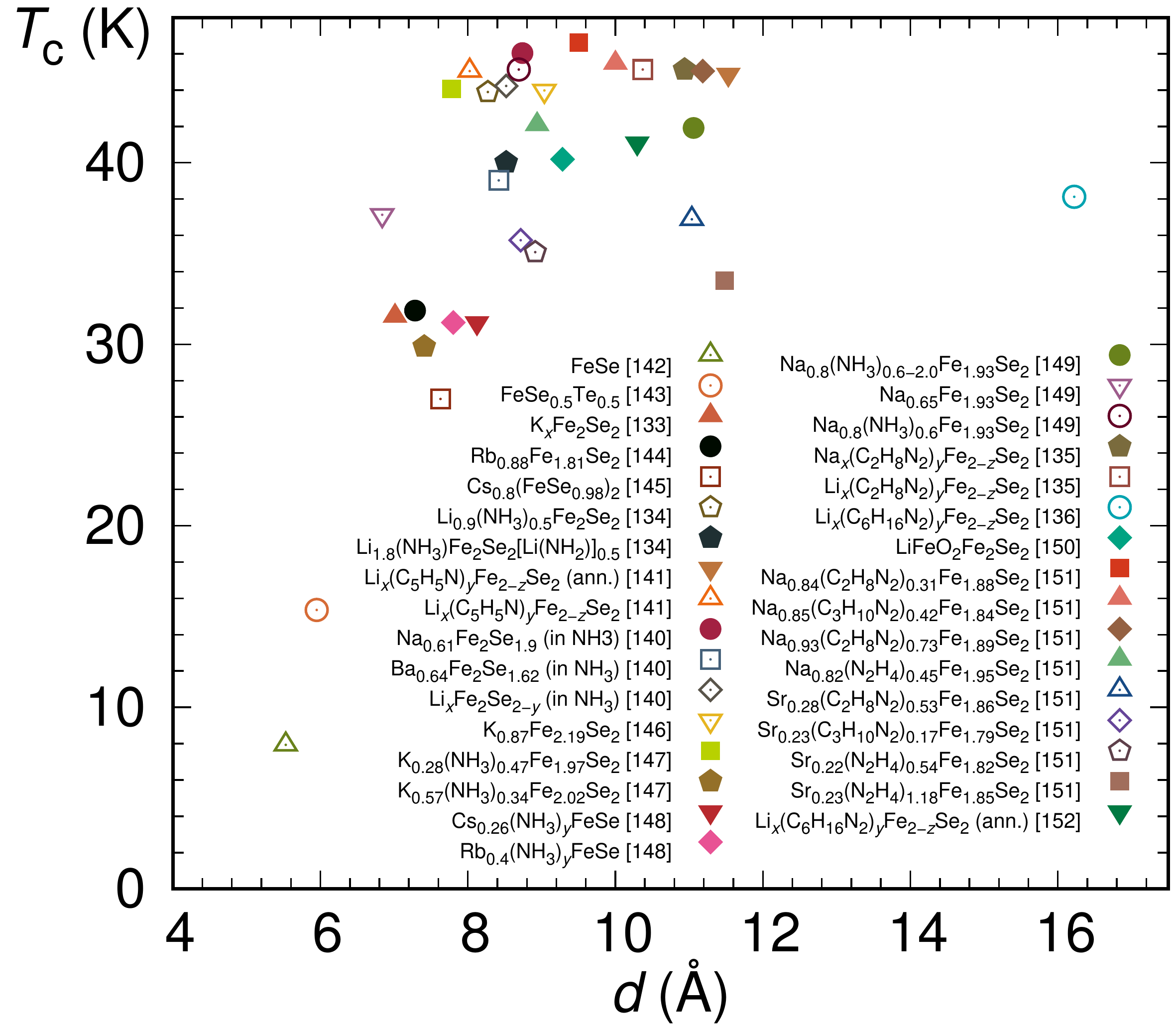}
\caption{Relationship between the maximum $T_c$ and the interlayer spacing in 
intercalated FeSe-based superconductors.}
\label{fig:Intercalate:tcplot}
\end{figure}

Another possibility to modify FeSe is electron doping. It was recently 
shown that various alkali atoms and organic molecules can be intercalated 
between the layers of bulk FeSe by different chemical 
processes~\cite{Guo2010,Scheidt2012,Noji2014,Hosono2014,Sedlmaier2014,
Burrard-Lucas2013, Sun2015,Ying2012, Krzton-Maziopa2012}. In this way, 
$T_c$ can be enhanced to up to $46~\mathrm{K}$ without application of pressure. 
Noji {\it et al.}~\cite{Noji2014, Hosono2014} correlated $T_c$ with enhanced 
interlayer spacing due to intercalation (see Fig.~\ref{fig:Intercalate:tcplot}
where data from 
Refs.~\onlinecite{Hsu2008,Yeh2008,Guo2010,Wang2011A,Krzton2011,Scheidt2012,Krzton-Maziopa2012,
Ying2012,Zhang2013,Ying2013,Zheng2013,Guo2014,Noji2014,Hosono2014,Lu2014,Hayashi2015,
Hosono2016} have been plotted), but could not 
explain the wide variation of transition temperatures found upon intercalation of 
different combinations of alkali atoms and organic solvents.

We performed DFT calculations for lithium und ammonia intercalated FeSe and 
showed that the initial rise of $T_c$ up to an interlayer spacing of about 
$9~\mathrm{\AA}$ can be explained with an increasingly two-dimensional 
electronic structure~\cite{Guterding2015a}. Beyond an interlayer distance of 
$9~\mathrm{\AA}$ the electronic structure is entirely two-dimensional and no 
further increase of $T_c$ can be expected through this mechanism. Using RPA 
calculations we furthermore find that the electron doping significantly 
modifies the superconducting pairing strength, and hence $T_c$, through a density 
of states effect as the upper edge of the hole bands moves closer to the Fermi 
level. Naturally this effect 
is also limited by the disappearance of hole pockets upon further electron 
doping.

Our predictions have been subsequently confirmed by various experimental groups, who 
could not enhance the transition temperatures by separating the FeSe layers 
further, but found a strong dependence of $T_c$ on the number of electrons 
doped~\cite{Guo2014, Hayashi2015, Yusenko2015, Hosono2016}. A large number of 
FeSe intercalates has been synthesized meanwhile, but $T_c$ empirically seems to 
be limited to $\leq 46~\mathrm{K}$. 

\section{Conclusions}
\label{sec:conclusions}
We reviewed the properties of iron-based superconductors under application of 
pressure and chemical doping. We discussed the influence of pressure on 
structural details of the 122 family of iron pnictides and in particular the 
emergence of the volume collapsed phase. Based on DFT+DMFT calculations we 
elucidated the issue of bad metallicity and its connection to Hund's rule 
coupling in a broad range of iron-based materials, in particular the hole-doped 
end members of the 122 series. Furthermore, we reviewed the nematic state and 
non-monotonous pressure dependence of superconductivity in bulk iron selenide. 
Finally we discussed the superconducting properties of a hole-doped 122 material 
and the intercalated iron selenides.

\section*{Acknowledgments}
The work presented in this review was made possible by the financial support 
provided by the German Research Foundation (Deutsche Forschungsgemeinschaft)  
through priority program SPP 1458. The authors thank Bernd B\"uchner, 
Rudi Hackl, Carsten Honerkamp and Dirk Johrendt for initiating this priority 
program and for important discussions.

\bibliography{valenti_group_publications,all_sources}

\end{document}